\documentclass[amsmath,amssymb,
onecolume,
superscriptaddress,12pt]{revtex4-2}

\usepackage[utf8]{inputenc}

\usepackage{lineno}

\usepackage{color}
\usepackage{graphicx}
\usepackage{amsmath}
\usepackage{float}
\usepackage{geometry}
\usepackage{indentfirst}
\usepackage[colorlinks=true,allcolors=blue]{hyperref}
\usepackage[english]{babel}
\begin{document}

\title{\textbf{Evolution of cooperation with joint liability}}
\author{Guocheng Wang}
\affiliation{\small Center for Systems and Control, College of Engineering, Peking University, Beijing 100871, China}
\author{Qi Su}
\affiliation{\small Center for Mathematical Biology, University of Pennsylvania, Philadelphia, PA 19104, USA}
\affiliation{\small Department of Mathematics, University of Pennsylvania, Philadelphia, PA 19104, USA}
\affiliation{\small Department of Biology, University of Pennsylvania, Philadelphia, PA 19104, USA}
\author{Long Wang}
 \email{longwang@pku.edu.cn}
\affiliation{\small Center for Systems and Control, College of Engineering, Peking University, Beijing 100871, China}
\affiliation{\small Center for Multi-Agent Research, Institute for Artificial Intelligence, Peking University, Beijing 100871, China}

\begin{abstract}
``Personal responsibility'', one of the basic principles of modern law, requires one to be responsible for what he did.
However, personal responsibility is far from the only norm ruling human interactions, especially in social and economic activities.
In many collective communities such as among enterprise colleagues and family members, one's personal interests are often bound to others' ---
once one member breaks the rule, a group of people have to bear the punishment or sanction.
Such a mechanism is termed ``joint liability".
Although many real-world cases have demonstrated that joint liability helps to maintain collective collaboration, a deep and systematic theoretical analysis on how and when joint liability promotes cooperation is lacking.  
Here we use evolutionary game theory to model an interacting system with joint liability, where one's losing credit could deteriorate the reputation of the whole group.
We provide the analytical condition to predict when cooperation evolves in the presence of joint liability, which is verified by simulations.
We also analytically prove that joint liability can greatly promote cooperation.
Our work stresses that  
joint liability is of great significance in promoting the current economic propensity.

\end{abstract}

\maketitle

\section{Introduction}
``Personal responsibility'' is one of the basic principles of modern law and stresses that one must bear the responsibility for what he did \cite{Tyler2006,Dan-Cohen1992}. 
Despite its reasonability and prevalence in human society, it is far from the only norm ruling human social interactions and economic activities.
In fact, joint liability, which closely binds one's personal interests to others', abounds in collective communities, such as among enterprise colleagues, family members, friends, and even in primates society \cite{Kornhauser1994,Ren1997,Alchian1972,Clutton-Brock1995,Allport1954,Akerlof1970}. 
The case of Grameen Bank is the most representative example of joint liability.
Due to the lack of valuable collateral, the poor often have less chance to receive loans from the bank.
Muhammad Yunus, the founder of the Grameen Bank, introduced the mechanism of joint liability --- to get the loan, each borrower must have at least four guarantors (could be other poor); while once a poor people fail to pay the debt back, all including the guarantors will lose the credit and chance to get loans any more \cite{Hossain1988,Ghatak1999}.
Other examples include: misconduct of a fraction of foreigners will leave the natives with bad impressions of all foreigners \cite{Allport1954};
the poor quantity of some products in a shop leads to the trust deficit to all other products \cite{Akerlof1970}.
Besides human societies, joint liability widely exists in various animal communities. 
For example, after food sources being looted by other vervet monkeys, adult females would attack looters' relatives \cite{Clutton-Brock1995}. In the macaque society, members of different matrilineal groups form alliance, and once being attacked, they often retaliate against a vulnerable member of the attacker's group \cite{Clutton-Brock1995}.
Simply speaking, when a group of people share the joint liability, once one breaks the rule, all of the rest have to bear the punishment or sanction.

In the example of Grameen Bank, after the poor get the loan from the bank, to maximize their temporary interests, they are expected to refuse to pay the debt back.
From the perspective of evolution and competition, all guarantors and the borrower lose the credit, while the borrower owns the extra benefit of not paying the debt back.
Therefore, theoretically, all individuals eventually evolve into credit-losers.
However, the history has seen the success of joint liability mechanism in promoting the prosperity in economic activities. 
How does joint liability affect individuals' decision-making?
This counter-intuition actually raises a long-standing question in evolutionary biology --- given one's instinct to increase his own benefits, why would cooperation (a kind of altruistic behavior, paying a cost to benefit strangers) prevail?
The past decades have seen numerous studies into this cooperation conundrum, and many insightful mechanisms have been revealed, such as direct reciprocity \cite{Hilbe2017,Nowak1993}, indirect reciprocity \cite{Nowak2005,Hilbe2018,Nowak1998,Santos2018}, spatial reciprocity \cite{Ohtsuki2006,Su2019a,Nowak1992,Allen2017,McAvoy2020}, and costly punishment \cite{Hauert2007,Helbing2010,Szolnoki2017,Henrich2006,Ohtsuki2009}.
In the case of costly punishment, besides cooperators and non-cooperators (called ``defectors"), the third type of individuals, called ``punishers", is introduced.
Punishers behave like cooperators and take the altruistic behavior, while they also enforce a punishment $\beta$ to the defector at a cost of $\gamma$.
In fact, the introduction of punishers leads to a ``second-order" social dilemma --- compared with cooperators, they pay an extra cost to punish defectors and thus put themselves at a disadvantageous place.
Therefore, the punishment mechanism does not certainly work to maintain a cooperative society \cite{Hauert2007,Colman2006,Ohtsuki2009}.

The success of Grameen Bank inspires us that joint liability might be facilitative to the evolution of cooperation.
Here we use evolutionary game theory to model an evolving system and introduce both the costly punishment and joint liability.
We derive a mathematical formula to predict the evolution of cooperation, which well agrees with the result obtained by simulations.
In particular, in the absence of joint liability, cooperation evolves only if $\beta-\gamma>4c$, where $c$ is the cost of taking cooperation.
When joint liability is present, the condition is relaxed to $\beta>\gamma$.
In other words, as long as the punishment $\beta$ exceeds the cost of punishment $\gamma$, cooperation evolves.
This simple rule supports that joint liability is of great significance in shaping the collective cooperation.

\section{Model}
Inspired by the example of bank loans, we propose a model of joint liability as follows.
The system consists of $N$ borrowers and one lender (e.g. a bank). 
The lender interacts with the borrower and decides whether or not to provide a loan $c$.
The borrower, if received the loan, makes an investment and derives a return $b$ ($b>c>0$).
For simplicity, we consider three types of borrowers, namely, punisher, cooperator, and defector.
The punishers and cooperators return the loan $c$ after deriving benefits, which gives them a net benefit $b-c$.
Defectors, however, choose to keep the loan and have a benefit $b$.

In each generation, the lender interacts with all borrowers, one by one.
In general, the lender has a tolerability of $m$ ($1\le m\le N$) --- after encountering $m$ defectors, the lender turns to reject lending loans to the rest.
Here we assume that after receiving the loan, cooperators and punishers immediately make investments and then return the loan to the lender, and therefore the lender can tell the type of the borrower before the next interaction.
$m=1$ means the lowest tolerability --- once encountering a defective individual, the lender rejects lending money to the rest.
$m=N$ corresponds to the maximal tolerability, lending money to all borrowers.

\begin{figure}[t]
\centering
\includegraphics[scale=0.29]{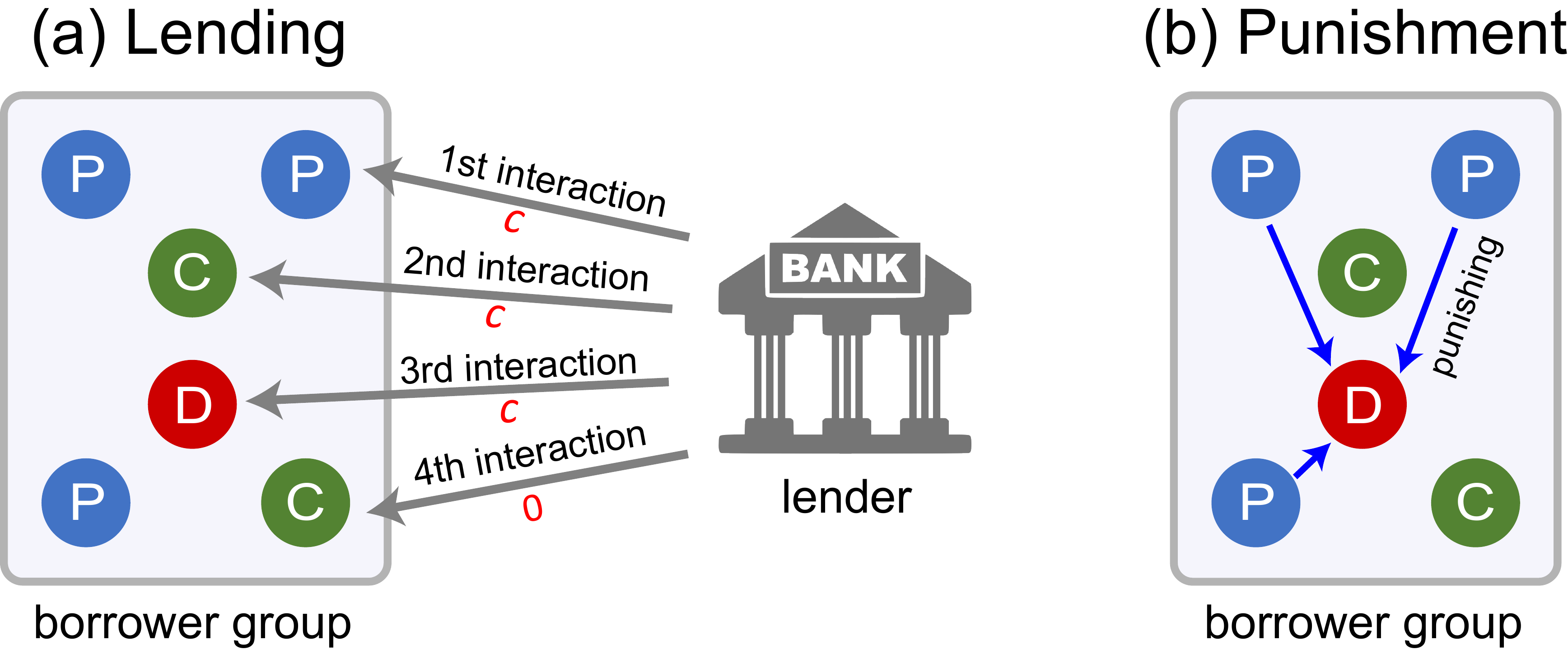}
\caption{\textbf{Illustration of joint liability.} 
We illustrate a system consisting of six borrowers and one lender with tolerability $m=1$.
Blue, green, and red circles represent punishers (P), cooperators (C), and defectors (D) respectively.
(a) Money lending stage. 
The lender interacts with all borrowers, one by one.
After encountering a defective borrower, the lender rejects lending money to the rest.
(b) Intragroup punishment stage.
Each punisher pays a cost $\gamma$ to impose every defector a fine $\beta$.}
\label{model_illustration}
\end{figure}

Following the loan lending stage, each punisher pays a cost $\gamma$ to impose a fine $\beta$ on each defector (Fig.~\ref{model_illustration}). To ensure that the total fine for a defector and the total cost for a punisher is bounded as the population size grows, we normalize $\gamma$ and $\beta$ by setting $\gamma/(N-1)$ and $\beta/(N-1)$. In the subsequent discussion, we call $\gamma$ punishing cost and $\beta$ fine.

Based on these settings, we can calculate the expected payoff for each borrower (see Methods for derivations). The payoffs for punisher, cooperator, and defector are respectively
\begin{align}
    \pi_P&=\min\left\{\frac{m}{N_D+1},1\right\}(b-c)-\frac{\gamma}{N-1}N_D,  \\ 
     \pi_C&=\min\left\{\frac{m}{N_D+1},1\right\}(b-c), \\ 
      \pi_D&=\min\left\{\frac{m}{N_D},1\right\}b-\frac{\beta}{N-1}N_P.
      \label{payoff}
\end{align}
Here $N$ is the population size, and $N_P$, $N_C$, $N_D$ are the numbers of punishers, cooperators, defectors respectively. We have $N_P+N_C+N_D=N$.

After both loan lending and intragroup punishment stages, borrowers update their strategies according to the classic pairwise comparison rule \cite{Nowak2006book}. 
Here we take account of both random strategy exploration and imitation of successful strategies. 
Specifically, a random borrower $x$ is selected.
With probability $\mu$, he switches to a random type, i.e. punisher, cooperator, or defector.
With probability $1-\mu$, another random borrower $y$ is selected among the rest.
$x$ compares her payoff ($\pi_x$) with $y$'s ($\pi_y$) and imitates $y$'s strategy with probability 
\begin{equation}
    \frac{1}{1+\exp{(-\kappa(\pi_y-\pi_x))}}.
\end{equation}
$\kappa$ denotes the selection strength.
A larger value of $\kappa$ means the increasing likelihood of imitating the individual with a higher payoff \cite{Wu2013}.

So far we have presented the minimal model, which helps to deliver the idea of joint liability.
We can further extend the model to capture more realistic interaction scenarios.
For example, besides the peer punishment, borrowers might interact with each other and defectors might enforce extra selection pressure on cooperators. 
In Section 3.3, we investigate the case that borrowers play donation game, where a cooperator/punisher pays a cost of $c'$ to bring his opponent a benefit $b'$, $b' > c' > 0$, while defectors do nothing.
The donation game is a classic Prisoner's Dilemma, where defection dominates cooperation \cite{Nowak2006}.
Also, we consider the case where only those defectors identified by the bank are punished.
We have demonstrated that all results are qualitatively consistent regardless of the model complexity.

\section{Results}
\subsection{Finite population}

We begin with the setup of rare mutation, in which the system is bounded to evolve into a homogeneous state (i.e. all-cooperator, all-defector, or all-punisher state) before the occurrence of a new mutant.
We calculate the fixation probability that the mutant of one type (cooperator, defector, or punisher) takes over the whole population of another type \cite{Nowak2004}.
The fixation probability measures the competition between two types, namely, one's ability to invade the other type. 
Using the method of Fudenberg and Imhof \cite{Fudenberg2006}, we study the average abundance of the three types throughout the evolutionary process.

\begin{figure}[t]
\centering
\includegraphics[scale=0.24]{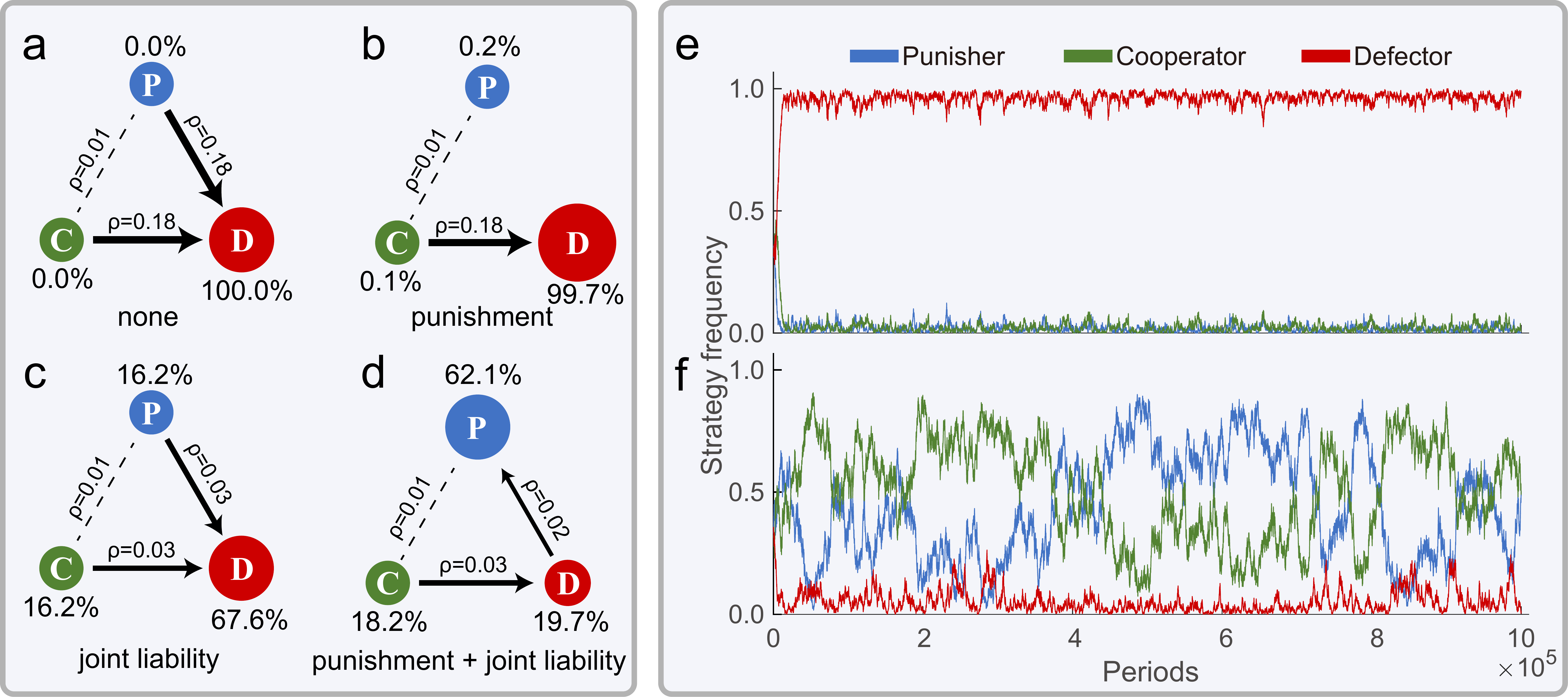}
\caption{\textbf{Joint liability boosts cooperation markedly.}  
For the sufficiently small mutation rate, the system is bounded to evolve into the homogeneous states, full punishers ($\textbf{P}$), full cooperators ($\textbf{C}$), or full defectors ($\textbf{D}$).
$\rho$ denotes the fixation probability that a single individual of one type takes over the whole population of the other type (only fixation probability exceeding $1/N$ is presented here).
The size of circles and values alongside represent the abundance of strategies.
(a) Case with neither joint liability nor punishment mechanism. Defectors dominate.
(b) Case with punishment while no joint liability. Punishment mechanism alone does not rescue cooperation.
(c) Case with joint liability while no punishment. Joint liability weakens the advantages of defectors over cooperators/punishers.
(d) Case with both joint liability and punishment mechanism. Cooperators/punishers dominate. 
(e) A simulated evolutionary process without joint liability, corresponding to (b). 
(f) A simulated evolutionary process with joint liability, corresponding to (d).
Parameters: $N=100$ (a--d), $N=300$ (e, f), $\kappa=0.2$, $b=3$, $c=1$, $\gamma=0.3$, $\beta=2$, $m=1$, and $\mu=0.01$ (e, f).}
\label{fig:simulation}
\end{figure}

We investigate four representative interaction situations, namely, one with neither punishment mechanism nor joint liability, one with either only punishment mechanism or only joint liability, and one with both of them. 
Fig.~\hyperref[fig:simulation]{2a} shows that without punishment mechanism and joint liability, defectors easily invade both populations of cooperators and of punishers, and they dominate the population across the evolutionary process.
The introduction of punishment mechanism leads to a slight fluctuation to the evolutionary outcome, but the effects are negligible (see Fig.~\hyperref[fig:simulation]{2b}).
Therefore the punishment mechanism alone does not rescue cooperation at all.
Intriguingly, joint liability alone is capable of maintaining a high level of cooperation and greatly weakens defectors' invasion to cooperators or punishers, highlighting more notable effects on cooperation relative to punishment mechanism (see Fig.~\hyperref[fig:simulation]{2c}).
We see that the combination of punishment mechanism and joint liability elevates the cooperation remarkably --- makes it possible for punishers to invade the population of defectors, and reduces defectors' abundance to lower than $20\%$ (see Fig.~\hyperref[fig:simulation]{2d}).
We further confirm the theoretical predictions by Monte Carlo simulations (see Fig.~\hyperref[fig:simulation]{2e, f}).

\begin{figure}[t]
    \centering
    \includegraphics[scale=0.5]{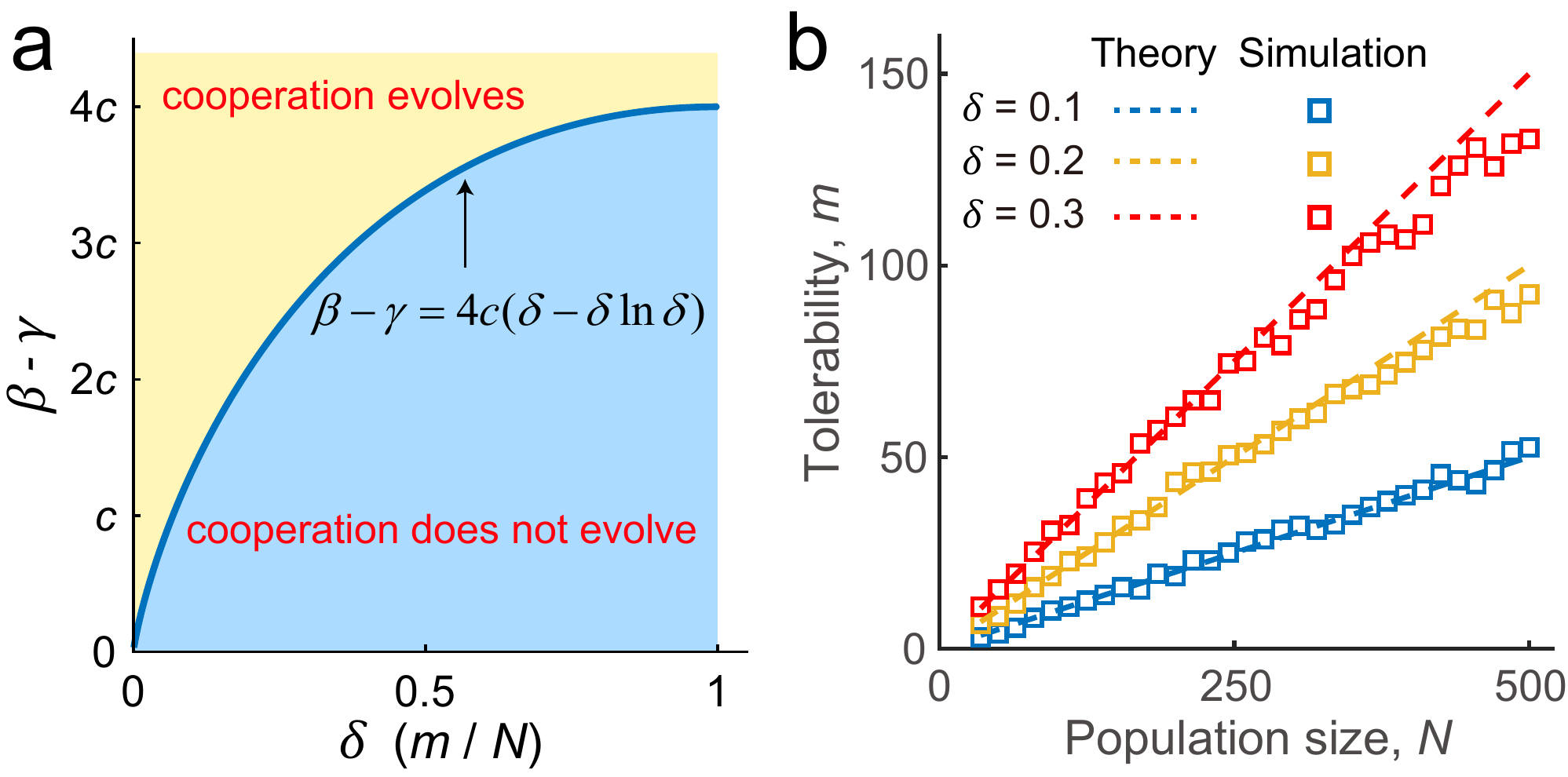}
    \caption{\textbf{The decreasing tolerability provides more benefits to cooperation.} 
    (a) Condition for the evolution of cooperation.
    The decreasing tolerability, i.e.  $\delta=m/N$, makes it possible for cooperation to evolve in a broader range of parameters, i.e. $\beta-\gamma$.
    (b) Results by simulations (colored squares) agree well with analytical predictions (dashed lines). We choose three sets of punishment parameters (i.e, $\beta-\gamma$): 1.32, 2.09, and 2.64. Under these settings, we perform simulations by searching for the value of $m$ with which the stationary abundance of defectors is close to $1/3$ (colored squares). On the other hand, the critical $\delta$ can be calculated by equation~(\ref{condition}) analytically, which are 0.1, 0.2, and 0.3 respectively (dashed lines).
    Parameters: $\mu=2.5\times 10^{-3}$, $s=0.04$, $b=3$, $c=1$, $\gamma=0.3$.
    }
    \label{fig:condition}
\end{figure}

Since both punishers and cooperators return the loans borrowed from the lender, they are into the category of altruists. 
Selection favors altruists over defectors only if the former's abundance exceeds $2/3$.
Here we derive the exact condition for the evolution of altruists (see Methods): under weak selection ($\kappa\to 0$) and in sufficiently large populations ($N\to \infty$), altruists are favored if and only if the fine to defectors, $\beta$, exceeds the punishing cost, $\gamma$, by a tolerability-based threshold, given by 
\begin{equation}
    \beta-\gamma>4c(\delta-\delta\ln{\delta}),
    \label{condition}
\end{equation}
where $\delta=m/N$, denoting the ratio of defectors the lender can tolerate. 
Function $\delta-\delta\ln{\delta}$ is monotone increasing (Fig.~\hyperref[fig:condition]{3a}). So, the more tolerant the lender is, the more difficult it is for altruistic behavior to evolve. 
For the extreme case with tolerability $m=1$ and a sufficiently large population size $N$, the condition for altruists to evolve is 
\begin{equation}
    \beta>\gamma.
    \label{condition1}
\end{equation}
Therefore, for sufficiently small tolerability, the evolutionary outcome is determined by fine $\beta$ and punishing cost $\gamma$, while independent of benefit $b$ and loan $c$.
The case of $m=N$ means the largest tolerability --- the lender lends to every individual even when all of them are defectors.
The largest tolerability in fact is equivalent to the absence of joint liability.
And we have the condition for the evolution of altruists, given by 
\begin{equation}
    \beta>\gamma+4c.
    \label{condition2}
\end{equation}
Equations~(\ref{condition1}, \ref{condition2}) also support that joint liability relaxes the threshold for altruists to evolve and thus promotes cooperation.
We further verify the theoretical analysis by Monte Carlo simulations.
As in Fig.~\hyperref[fig:condition]{3b}, we fix three sets of parameters of punishment (i.e.  $\beta-\gamma$) and calculate the respective critical tolerability $m$ below which the abundance of defectors is lower than $1/3$.
For different choices of population size $N$, our analysis (equation~(\ref{condition})) exactly predicts the evolutionary outcome.

\begin{figure}[t]
\centering
\includegraphics[scale=0.45]{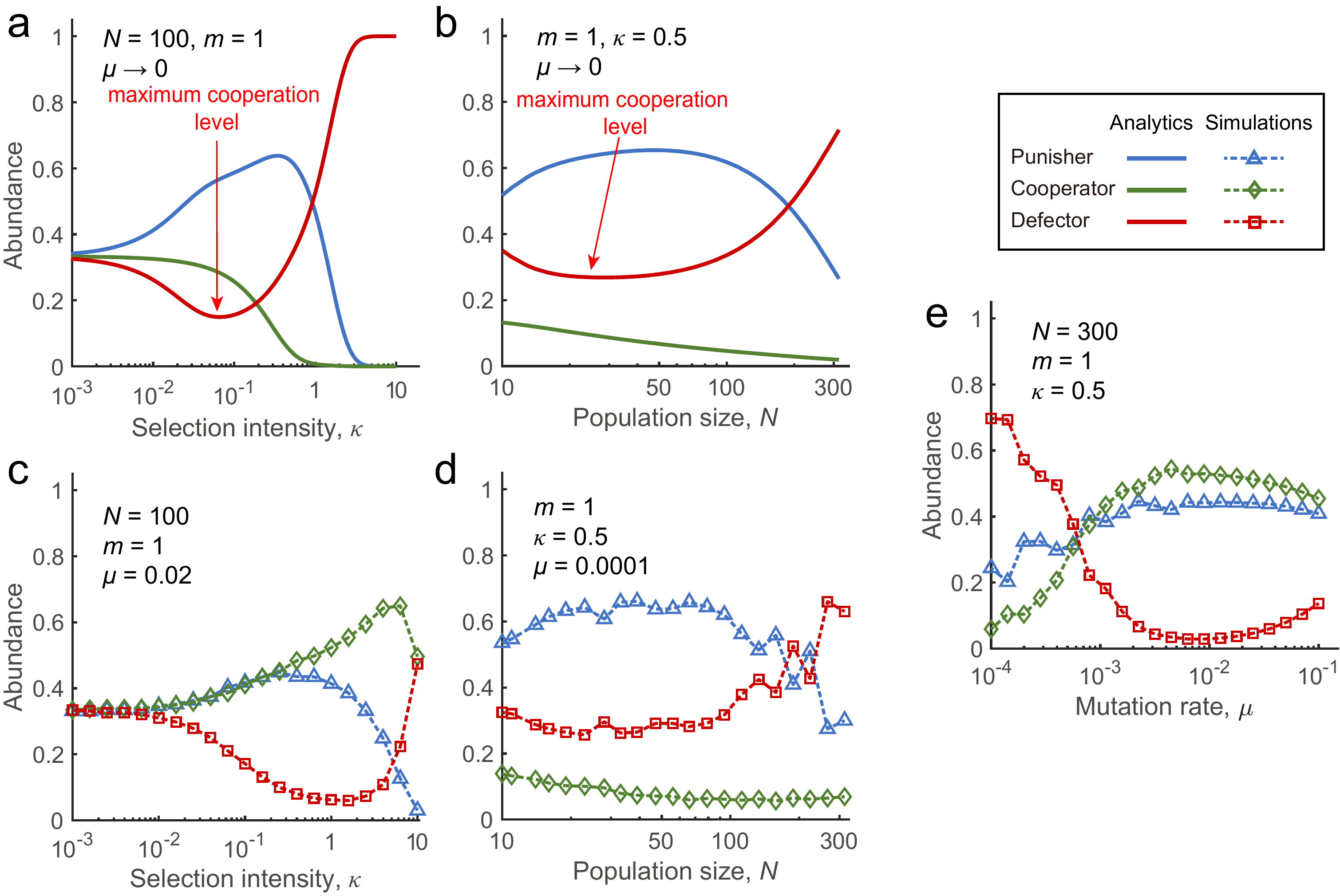}
\caption{\textbf{Medium-sized selection intensity, population size, and mutation rate benefit cooperation most.} We illustrate the abundance of three types of borrowers as a function of selection intensity $\kappa$ (a, c), population size $N$ (b, d), and mutation rate $\mu$ (e). 
Panels (a, b) present analytical results and (c, d, e) show results by individual-based simulations.
The analytical predictions qualitatively agree with the results of simulations, i.e. panels (a, c) and (b, d).
An intermediate level of selection intensity, population size, and mutation rate are most beneficial to cooperation.
Parameters: $b=3$, $c=1$, $\gamma=0.3$, $\beta=2$.}
\label{fig-abundance}
\end{figure}

We proceed with investigations into how selection intensity, population size, and mutation rate, affect the evolutionary dynamics.
We study a broad parameter range by both theoretical analysis and Monte Carlo simulations.
It turns out that the introduction of joint liability can maintain the advantages of cooperators and punishers over defectors for a wide range of selection intensity (Fig.~\hyperref[fig-abundance]{4a, c}).
An extremely small selection intensity leads to the evolutionary process close to that under neutral drift, while an extremely large value leads to a nearly deterministic process \cite{Wu2015}, which does not support the evolution of cooperation.
An intermediate value of selection intensity, therefore, sees the highest level of altruists, or say, the lowest level of defectors.
Analogously, an intermediate population size $N$ is most beneficial to the evolution of cooperators and punishers (Fig.~\hyperref[fig-abundance]{4b, d}).

For the increasing mutation rate $\mu$, the evolution sees an initial abundance increase in cooperators and punishers, and a following decrease (Fig.~\hyperref[fig-abundance]{4e}).
Note that extremely small mutation rates reduce the randomness of the evolving process, and extremely large mutation rates lead to a totally random system.
When the mutation rate lies between the two cases, the evolution often sees abundant cooperation. The optimality of intermediate selection intensity and mutation rate both indicate that appropriate randomness is of significance to foster cooperation.
Here we provide a few intuitions about the underlying mechanisms.
Since paying an extra cost to punish defectors, punishers put themselves at disadvantageous positions compared with cooperators.
The increasing selection intensity leads to the extinction of punishers.
Once punishers are eliminated, defectors are free of punishment and dominate the population. 
However, an intermediate selection intensity enables punishers to eliminate defectors first.
The evolution, therefore, ends up with the coexistence of punishers and cooperation since the equal payoff. 
Also, when the mutation rate is properly large, once the punishers become extinct, a new punisher mutant is likely to appear, which is adverse to defectors.
Therefore, a proper mutation rate decreases defectors' advantages throughout the evolutionary process.
The following study of infinite populations further confirms our analysis here based on a finite number of individuals.

\subsection{Infinite population}
\label{section:3.2}
The setup of an infinite population could lead to totally different evolutionary dynamics from a finite population \cite{Traulsen2005,Molina2021}. 
Here to prove the generality of the joint liability's cooperation-promoting effects, we study the infinite population by means of the classic replicator equations.
Let $x,y,z$ denote the proportions of punishers, cooperators, and defectors respectively, which gives $x+y+z=1$. According to the payoff expressions in the case of finite population, for $N \to \infty$, we obtain
\begin{align}
     \pi_P=&\min\left\{\frac{\delta}{z},1\right\}(b-c)-\gamma z,  \\
    \pi_C=&\min\left\{\frac{\delta}{z},1\right\}(b-c), \\
    \pi_D=&\min\left\{\frac{\delta}{z},1\right\}b-\beta x.
\end{align}
The average payoff is then given by: $\bar{\pi}=x\pi_P+y\pi_C+z\pi_D$. We assume that every player with strategy $i\in\{P,C,D\}$ adopts another randomly selected player's strategy, with a probability proportional to the difference between the two player's payoffs. Let $x_i$ denote the proportion of players who adopt strategy $i$ in the population. The replicator equation can be written as

\begin{equation}
    \dot{x}_i=x_i(\pi_i-\bar{\pi}).
\end{equation}
Substituting $\pi_P$, $\pi_C$, and $\pi_D$ into the replicator equation, we can obtain three differential equations. With the relation $x+y+z=1$, the system is actually governed by two independent variables and equations. Choosing $x$ and $z$ as the independent variables, the replicator equations can be simplified to

\begin{equation}
    \left\{
    \begin{aligned}
         \dot{x}=&x\left[(\gamma+\beta)xz-\gamma z-\min\left\{\frac{\delta}{z},1\right\}cz \right],  \\
         \dot{z}=&z\left[(\gamma+\beta)xz-\beta x-\min\left\{\frac{\delta}{z},1\right\}(cz-c) \right]. 
    \end{aligned}
    \right.
\label{replicator_equation}
\end{equation}

\begin{figure}
    \centering
    \includegraphics[scale=0.8]{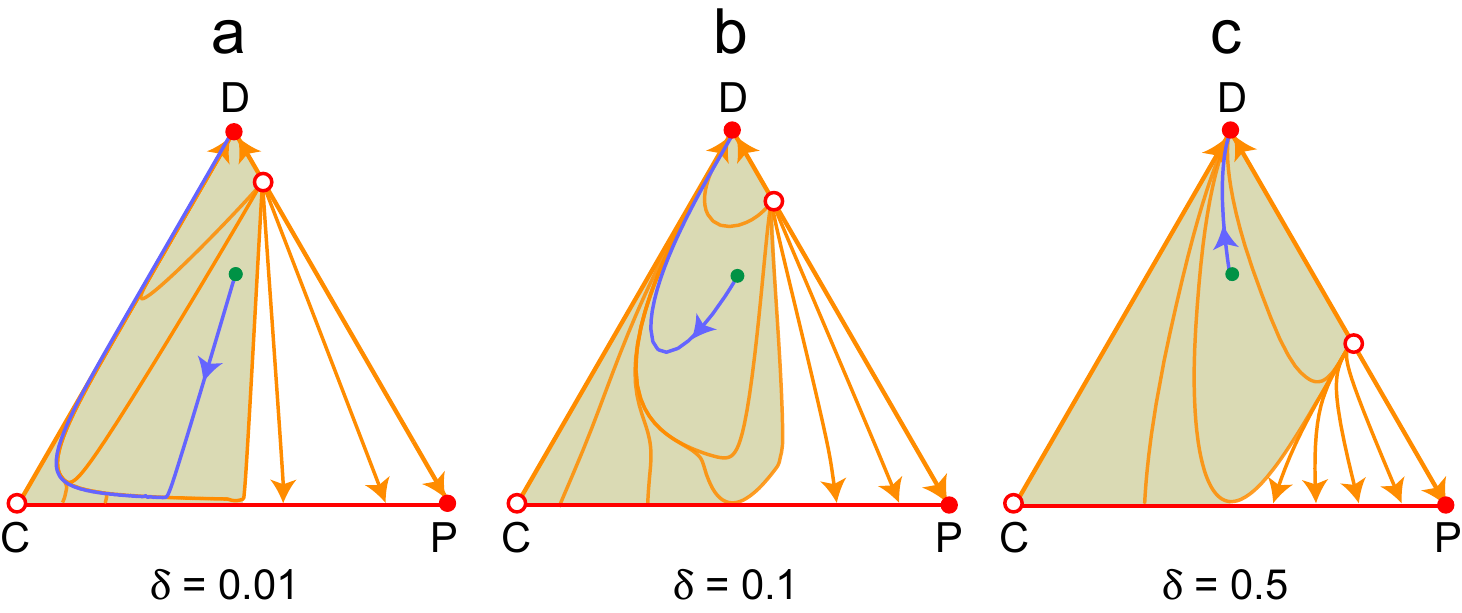}
    \caption{\textbf{The cooperation-promoting effects of joint liability in infinite populations.} Let $e=(x,y,z)$ denote the frequency of punishers, cooperators, and defectors, with $x+y+z=1$. The triangles represent the frequencies of the three types of individuals. Each arrow line represents a trajectory. The equilibrium points are marked in red. The dark green region is the attraction domain of the state where all individuals are defectors, and the rest region is the attraction domain of the invariant manifold $\chi=\{e=(x,1-x,0)|c/\beta<x<1\}$. All points on manifold $\chi$ are cooperative states. As tolerability $\delta$ decreases, the attraction domain of $\chi$ becomes larger. It means small $\delta$ can boost cooperation. Even though in the dark green region, trajectories do not converge to cooperative states, small $\delta$ can also change their directions and make them close to cooperative states. For example, let the trajectories start from the same point (the green point) in (a), (b), and (c),  the smaller $\delta$ is, the closer the trajectory will get to the line $z=0$. Parameter settings: $\beta=2$, $\gamma=0.3$, $c=1$.}
    \label{simplex}
\end{figure}

We then analyze equilibrium points of the above equation, which is obtained by taking the right-hand side of equation~(\ref{replicator_equation}) to be $0$. We denote the equilibrium point by $e=(x^*,y^*,z^*)$. There are five solutions: four points\\
 \begin{tabular}{ll}
         $e_1=(0,0,1)$, & $e_2=\left(\frac{2\gamma+\beta-\sqrt{\beta^2-4\delta c(\gamma+\beta)}}{2(\gamma+\beta)},0,\frac{\beta+\sqrt{\beta^2-4\delta c(\gamma+\beta)}}{2(\gamma+\beta)}\right)$, \\
         $e_3=\left(\frac{2\gamma+\beta+\sqrt{\beta^2-4\delta c(\gamma+\beta)}}{2(\gamma+\beta)},0,\frac{\beta-\sqrt{\beta^2-4\delta c(\gamma+\beta)}}{2(\gamma+\beta)}\right)$, & $e_4=(\frac{\gamma+c}{\gamma+\beta},0,\frac{\beta-c}{\gamma+\beta})$,
    \end{tabular}\\
and a line $l:z=0$. $e_1$ and $l$ always exist. However, $e_2$ and $e_3$ exist only when their third elements $z^*$ satisfy $z^*>\delta$; $e_4$ exists only when its element $z^*$ satisfies $z^*<\delta$. In Methods, we present all derivative details.

Furthermore, we use the method of linearization to analyze the stability of each equilibrium solution (see Methods). 
All conditionally existing equilibrium points (i.e. $e_2$, $e_3$, and $e_4$) are unstable. Only $e_1$ and a subset of $l$ are stable. Specifically, only the points in $\chi=\{e=(x,1-x,0)|c/\beta<x\le 1\} \subset l$ are stable. 
For a better understanding, we use simplex to illustrate the trajectories (see Fig.~\ref{simplex}). 
Throughout the evolutionary process, all trajectories starting from the interior of the simplex converge to $e_1$ or $\chi$. The point $e_1$ represents the state where all individuals are defectors, and the set of points on $\chi$ means the population consist of only punishers and cooperators. We call the attraction basin of $e_1$ ($\chi$) defective region (cooperative region).

Fig.~\ref{simplex} shows that the decreasing tolerance ratio $\delta$ is more beneficial to cooperation, which can be understood from two perspectives.
Firstly, the cooperative region enlarges as the tolerance ratio $\delta$ decreases.
Therefore, when beginning with a random initial configuration, the system is increasingly likely to evolve into the cooperator-punisher state.
In addition, the decreasing $\delta$ can change the evolutionary trajectories.
For example, although the blue trajectories in Fig.~\hyperref[simplex]{5a, b, c} have the same initial configuration (green dots) and the stationary state (i.e., $e_1$), the evolutionary trajectories are remarkably different.
The trajectory for small $\delta$ (see Fig.~\hyperref[simplex]{5a}) could be greatly close to the cooperator-punisher state (i.e., $l$) during the evolutionary process.
With any noise or perturbation, the trajectory might fluctuate to the cooperator-punisher state and then become fixed, which is easier than those far away from the cooperator-punisher state.
Once again, randomness is expected to play a critical role for joint liability to promote cooperation.

\subsection{Intragroup interactions}
\begin{figure}[t]
    \centering
    \includegraphics[scale=0.29]{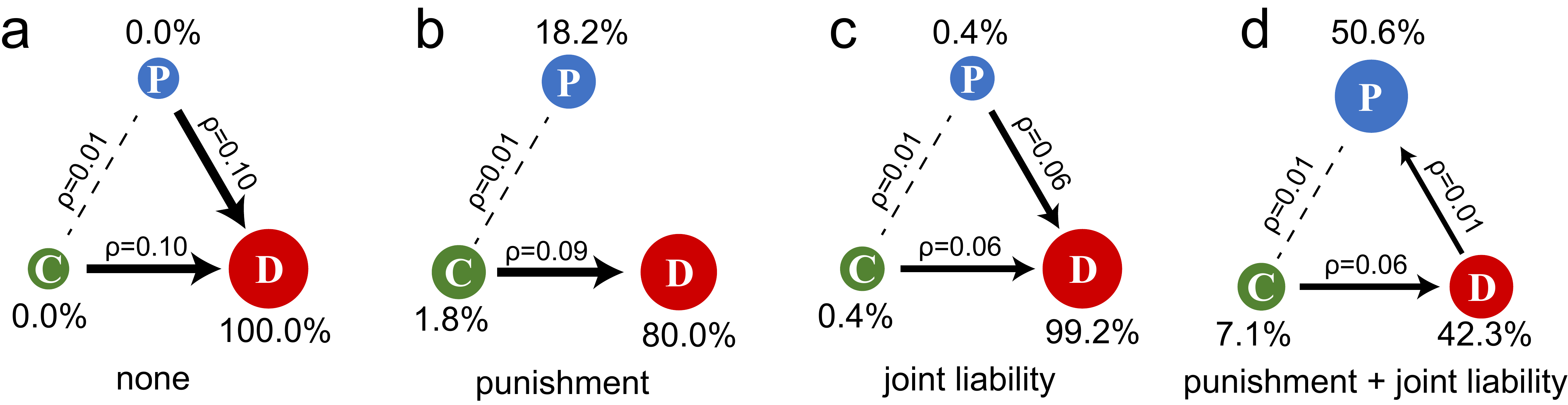}
    \caption{\textbf{Joint liability stabilizes cooperation when intragroup interactions are considered.} 
    Here all intragroup borrowers play donation games.
    We investigate four cases, with neither punishment and joint liability (a), with punishment alone (b), with joint liability alone (c), with both punishment and joint liability (d).
    Punishment or joint liability alone could slightly increase cooperation. 
    And they together can elevate cooperation to a remarkably high level. 
    Parameters: $N=100$, $m=1$, $s=0.05$, $\gamma=0.5$, $\beta=6$, $b=b'=3$, $c=c'=1$.}
    \label{fig:inner interaction}
\end{figure}
Finally, we extend the research by enabling the 
intragroup interactions --- individuals within the lending group interact with each other and derive payoffs.
We assume they play donation games (other games can be studied analogously), a classic prisoner's dilemma game where defection dominates.
In a donation game, a cooperator/punisher offers a benefit $b'$ to his opponent and incurs a cost $c'$; a defector pays nothing.
Individuals' payoff is given by
\begin{align}
    \pi_P=&\min\left\{\frac{m}{N_D+1},1\right\}(b-c)-\gamma\frac{N_D}{N-1}+\frac{N-N_D-1}{N-1}b'-c', \\
    \pi_C=&\min\left\{\frac{m}{N_D+1},1\right\}(b-c)+\frac{N-N_D-1}{N-1}b'-c',\\
    \pi_D=&\min\left\{\frac{m}{N_D},1\right\}b-\beta\frac{N_P}{N-1}+\frac{N-N_D}{N-1}b'.
\end{align}

The analogous analysis into four cases (Fig.~\ref{fig:inner interaction}) further confirms that 
cooperation thrives only when both punishment and joint liability work together.
The explicit condition for the evolution of cooperation with intragroup interactions is
\begin{equation}
    \beta-\gamma>4c(\delta-\delta \ln \delta)+4c'.
\end{equation}

\section{Discussion}
How cooperators evolve from a population of defectors has long been the enigma in the field of evolution, and has received decades of studies by researchers from different disciplines  \cite{Nowak2006,Ohtsuki2006,Traulsen2006,Antal2009,Constable2016,Hauert2007,Chen2015inter,Su2019a}. However, nearly all prior studies have assumed that every individual, regardless of 
taking strategic behavior in a social context or presenting phenotype in a genetic context, acts and evolves in an isolated way.
Here we, for the first time, introduce the mechanism of joint liability that binds one's interests with others'. Given the prevalence in human's daily and economic activities, our work is of great significance in theoretically evidencing the positive effects of joint liability on long-term societal prosperity.

Indeed, in well-developed countries or regions, the 
wide applications of advanced technologies such as social media make one's reputation more visible to the public, and banks therefore can easily make a decision whether or not to lend the loan by evaluating one's personal credit. 
But in communities lacking these public information, it is unlikely to accurately evaluate a specific person \cite{Posner1983}.
A feasible and effective approach is to refer to the reputation of the whole community.
When a sufficiently large proportion of members have low credits, the rest are also categorized into the class of credit-losers.
Banks, despite the lack of borrowers' information, can make use of the joint liability to mitigate the risks of lending loans to those with low credits.

We observe that the introduction of joint liability alone is capable of rescuing cooperation, which can never be achieved by the mechanism of punishment alone. 
Especially, the prior studies have shown that costly punishment can hardly evolve \cite{Hauert2007,Colman2006}, since it requires the punishers to pay an extra cost of implementing punishment, which thus reduces the punishers' advantages.
We also find that the combination of joint liability and the punishment mechanism contributes to cooperation most and boosts cooperation to a remarkably high level.
In fact, besides the costly punishment, many other mechanisms such as deporting defectors out of the group, once combined with joint liability, also serves as a great promoter of cooperation. 
We are left to conclude that the joint liability gives individuals incentives to punish defectors and the punitive measure offers individuals tools, which turns out to be indispensable in maintain the large-scale cooperation.

Furthermore, by both extensive simulations and theoretical analysis, we prove that the fewer defectors the bank tolerates, the more beneficial it is for cooperation to evolve. 
Due to the cooperation-promoting effects of joint liability, one might take the monotonous effect for granted.
However, we argue that although the smaller tolerability means less chance for defectors to cheat the banks (weakening the advantages of defectors), it also makes it harder for cooperators to increase their benefits (decreasing the interests of cooperators).
The evolutionary outcome actually depends on the balance between the influence on defectors and cooperators.
The evolution has seen the more detrimental effects to defectors and therefore the relative advantages to cooperators.

Although there are some prior studies about implicated punishment and reputation mechanism \cite{Chen2015,Hilbe2018,Fu2008,Suzuki2005,DosSantos2011}, we stress that models therein are different from ours. The implicated punishment assumes that an external supervisor directly imposes a fine on all players when there are defectors in the group \cite{Chen2015}. And prior studies about reputation mechanism consider personal reputation --- once taking defection, one loses his reputation, and then all other players cease to cooperate with him. The effectiveness of our model, however, relies on collective reputation ---
one's personal behavior can affect the whole group’s reputation, which in turn affects the personal reputation of other members'. A player affiliated to a group with good reputation can obtain an identification of cooperation even though he never shows the preference to cooperate before. Examples of collective reputation are ubiquitous, such as a product’s brand and a student’s diploma, both serving as a credit certification of their qualities and abilities. 
Besides, to some degree, joint liability is analogous to the threshold public goods game, in which the total contribution into the common pool must exceed a threshold such that each individual can benefit from the public goods \cite{Pacheco2009,Su2019plos,Li2016}. In other words, if the number of defectors who do not contribute is above a certain threshold, none of them obtains payoff. In the joint liability model, the threshold is the lender's tolerability. If the number of defectors he encounters exceeds this threshold, all other players lose the opportunities to obtain benefits.

To the best of our knowledge, the current work is the first systematic investigation into how joint liability affects the evolution of prosocial behaviors. 
Given the prevalence of joint liability in all walks of life, a variety of realistic scenarios can be studied.
For instance, human society consists of communities of different cultures and religions.
Often, one's impression of a community is based on the limited information and survey to a small number of members, or one's behavior when interacting with a specific member is determined by his impression of the whole community. A promising research direction is to model multiple communities with joint liability.
Furthermore, a prior study has considered the case where the observation of one's reputation may be subject to noise \cite{Hilbe2018}, and find this observation error is vital for the evolutionary stability. Here, an extension worthy of studying is that the lender may judge the borrower by mistake. 
An occasional error in judging a cooperator may lead to all borrowers' losing opportunities of leading loans, which may have a great impact on the evolving process.
The realistic population is often structured and a large amount of literature has proved that the population structure can notably influence the evolutionary outcome \cite{Nowak1992, Ohtsuki2006, Allen2017, Su2019prsb, Allen2019, Li2020}.
Analogously, the population structure might play an important role in the establishment of the joint liability, i.e. only those socially connected sharing joint liability. 
Our work, here, presents the initial efforts, which shows joint liability is powerful in promoting cooperation. We expect many deeper insights along this direction.

\section{Methods}
\subsection{Payoff calculation}
Here, we calculate the expected payoffs for the three types of players. Suppose the lender's tolerability is $m$, which means the lender will reject lending to any other individuals in the group after encountering $m$ defectors. The group has $N$ individuals consisting of $N_P$ punishers, $N_C$ cooperators, and $N_D$ defectors. 

We first assume $N_D\ge m$. Let random variable $X$ denotes the number of individuals that the lender interacts with before he rejects lending. We can obtain the probability distribution of $X$ by making an analogy to the Polya urns problem: In an urn consisting of $N_C+N_P$ red balls and $N_D$ black balls, we extract balls one by one without replacement and stop extracting after we have extract $m$ black balls. Thus, the distribution of $X$ is
\begin{equation}
    \text{Prob}(X=k)=\frac{(N-k)!\begin{pmatrix}A\\k-m\end{pmatrix}(k-m)!\begin{pmatrix}k-1\\m-1\end{pmatrix}(N-A)!}{N!(N-A-m)!},
\end{equation}
where $A=N_P+N_C$ ($A+m\le N$). The lender provides the group's all cooperators and punishers with a payoff
\begin{equation}
    \sum_{k=m}^{A+m}\text{Prob}(X=k)[(k-m)(b-c)]=\frac{mA(b-c)}{N_D+1}
\end{equation}
in total. So, each cooperator and punisher can get a payoff $\frac{m(b-c)}{N_D+1}$. 

For defectors, the lender always lends to $m$ defectors in the group before he rejects lending. So, the lender offers all defectors $mb$ in total. Each defector can get a payoff $\frac{mb}{N_D}$ in average. 

When $N_D< m$, all individuals in the group can get loans. Thus, each punisher and cooperator obtains $b-c$ and each defector obtains $b$.

Considering each punisher imposes a fine $\beta/(N-1)$ on each defector at a cost of $\gamma/(N-1)$, we can finally obtain the payoffs for the three types of individuals: 
\begin{align}
    \pi_P&=\min\left\{\frac{m}{N_D+1},1\right\}(b-c)-\frac{\gamma}{N-1}N_D,  \\ 
     \pi_C&=\min\left\{\frac{m}{N_D+1},1\right\}(b-c), \\ 
      \pi_D&=\min\left\{\frac{m}{N_D},1\right\}b-\frac{\beta}{N-1}N_P.
\end{align}

\subsection{Fixation probability approximation}
If mutation is rare, a strategy's frequency at equilibrium can be calculated by computing the stationary distribution of a Markov chain \cite{Fudenberg2006}. This Markov chain has three states, which are the three homogeneous population states consisting entirely of punishers (\textbf{P}), cooperators (\textbf{C}), and defectors (\textbf{D}). The transition probability matrix $\mathbf{M}$ between these three states is
\begin{equation}
    \mathbf{M}=\begin{bmatrix}
    1-\mu\rho_{PC}/3-\mu\rho_{PD}/3 & \mu\rho_{PC}/3 & \mu\rho_{PD}/3 \\
    \mu\rho_{CP}/3 & 1-\mu\rho_{CP}/3-\mu\rho_{CD}/3 & \mu\rho_{CD}/3 \\
    \mu\rho_{DP}/3 & \mu\rho_{DC}/3 & 1-\mu\rho_{DP}/3-\mu\rho_{DC}/3
    \end{bmatrix}.
\end{equation}
Here, $\mu$ is the mutation rate, which does not affect the stationary distribution. Thus, we can set $\mu=3$. The element $\rho_{XY}$ denotes the probability that a single $Y$ individual can invade and take over the population consisting entirely of $X$ individuals.

If defectors are absent, cooperators and punishers have the same payoff, and the evolving process becomes a neutral drift. Thus, $\rho_{PC}$ and $\rho_{CP}$ both equal $1/N$. We now show how to calculate $\rho_{PD}$ as an example. When computing $\rho_{PD}$, we should assume that the population consist entirely of punishers and defectors. Then, a prior study \cite{Nowak2004} gives the expression of fixation probability, that is
\begin{equation}
    \rho_{PD}=\frac{1}{1+\sum\limits_{k=1}^{N-1}\prod\limits_{i=1}^{k}\frac{T_{PD}^{-}(i)}{T_{PD}^{+}(i)}}.
    \label{fixation_eq}
\end{equation}
Here, $T_{PD}^+(i)$ and $T_{PD}^-(i)$ are the probabilities that the number of defectors increases or decreases by one when there are $i$ defectors in the population. For the pairwise comparison rule, we have
\begin{equation}
    T_{PD}^+(i)=\frac{N-i}{N}\frac{i}{N-1}\frac{1}{1+\exp{(\kappa(\pi_P(i)-\pi_D(i)))}},
\end{equation}
\begin{equation}
    T_{PD}^-(i)=\frac{i}{N}\frac{N-i}{N-1}\frac{1}{1+\exp{(\kappa(\pi_D(i)-\pi_P(i)))}}.
\end{equation}
To obtain the punisher and cooperator's payoffs, we stress that if the lender's tolerability exceeds the number of defectors among the group, he will lend to all individuals in the group. So, the payoffs are
\begin{equation}
    \pi_P(i)=\left\{
    \begin{array}{ll}
         \frac{m(b-c)}{i+1}-\gamma \frac{i}{N-1} & \text{if }i\ge m  \\
         b-c-\gamma \frac{i}{N-1} & \text{if }i\le m-1
    \end{array}, \right.
\end{equation}
\begin{equation}
    \pi_D(i)=\left\{ 
    \begin{array}{ll}
        \frac{mb}{i}-\beta \frac{N-i}{N-1} & \text{if }i\ge m  \\
        b-\beta \frac{N-i}{N-1} & \text{if }i\le m-1
    \end{array}.
    \right.
\end{equation}

Substituting $T_{PD}^+$ and $T_{PD}^-$ into equation~(\ref{fixation_eq}), we obtain 
\begin{equation}
\begin{split}
     \rho_{PD}=&\frac{1}{1+\sum\limits_{k=1}^{N-1}\prod\limits_{i=1}^{k}\frac{T_{PD}^{-}(i)}{T_{PD}^{+}(i)}}  
    =\frac{1}{1+\sum\limits_{k=1}^{N-1}\prod\limits_{i=1}^{k}\frac{1+\exp{(\kappa(\pi_P(i)-\pi_D(i)))}}{1+\exp{(\kappa(\pi_D(i)-\pi_P(i)))}}}  \\
    =&\frac{1}{N+\kappa\sum\limits_{k=1}^{N-1}\sum\limits_{i=1}^{k}(\pi_P(i)-\pi_D(i))+o(\kappa)} \\
    =&\frac{1}{N+\kappa\sum\limits_{k=1}^{N-1}(N-k)(\pi_P(k)-\pi_D(k))+o(\kappa)}. 
\end{split}
\end{equation}
Remembering that $\sum_{i=1}^N \frac{1}{i} \approx \ln{(N+1)}+\zeta$, where $\zeta$ is the Euler's constant, we have
\begin{equation}
    \rho_{PD} \approx \frac{1}{N}\left[1-\kappa\left(-\frac{N}{6}\gamma+\frac{N}{3}\beta -\frac{m^2}{2N}c-mc\ln{(\frac{N}{m})}\right)\right]
    \label{eq:rho_pd}
\end{equation}
when $N\kappa \to 0$. Similarly, we can obtain $\rho_{DP}$, $\rho_{DC}$, and $\rho_{CD}$:
\begin{equation}
    \rho_{DP} \approx \frac{1}{N}\left[1-\kappa\left(\frac{N}{3}\gamma-\frac{N}{6}\beta+mc-\frac{m^2}{2N}c \right) \right],
    \label{eq:rho_dp}
\end{equation}
\begin{equation}
    \rho_{CD} \approx \frac{1}{N} \left[1-\kappa\left(-\frac{m^2}{2N}c-mc\ln{(\frac{N}{m})}\right)\right],
    \label{eq:rho_cd}
\end{equation}
\begin{equation}
    \rho_{DC} \approx \frac{1}{N} \left[1-\kappa\left(mc-\frac{m^2}{2N}c \right)\right].
    \label{eq:rho_dc}
\end{equation}

By computing the stationary distribution of $\mathbf{M}$, we can get the abundance of the three strategies. The abundance distribution is a row vector $\mathbf{v}$ satisfying $\mathbf{vM=v}$. Then, we can obtain that $\mathbf{v}=[v_P,v_C,v_D]/(v_P+v_C+v_D)$, where
\begin{equation}
    v_P=\rho_{CP}\rho_{DP}+\rho_{CP}\rho_{DC}+\rho_{DP}\rho_{CD},
\end{equation}
\begin{equation}
    v_C=\rho_{DP}\rho_{PC}+\rho_{PC}\rho_{DC}+\rho_{DC}\rho_{PD},
\end{equation}
\begin{equation}
    v_D=\rho_{CP}\rho_{PD}+\rho_{PC}\rho_{CD}+\rho_{PD}\rho_{CD}.
\end{equation}
If the frequency of defectors at equilibrium is less than $1/3$, we say natural selection supports the evolution of cooperation. Thus we obtain the condition for the evolution of cooperation:
\begin{equation}
    2v_D<v_P+v_C.
    \label{inequality_condition}
\end{equation}
Substituting all fixation probabilities (equations~(\ref{eq:rho_pd}, \ref{eq:rho_dp}, \ref{eq:rho_cd} and \ref{eq:rho_dc})) into equation~(\ref{inequality_condition}), we have
\begin{equation}
    \beta-\gamma-4\frac{m}{N}c+4c\frac{m}{N}\ln\frac{N}{m}>0.
\end{equation}
We call $m/N$ tolerance ratio, denoted by $\delta$. So the condition for the evolution of cooperation can also be written as
\begin{equation}
    \beta-\gamma-4\delta c+4c\delta \ln \delta>0.
\end{equation}

\subsection{Replicator dynamics analysis}
In Sec. \ref{section:3.2}, we have derived the replicator equations. Letting $x$ and $z$ be independent variables, the replicator equations are
\begin{equation}
	\label{replicator system}
\left\{
\begin{aligned}
         \dot{x}&=x[(\gamma+\beta)xz-\gamma z-\min\{\frac{\delta}{z},1\}cz ]  \\
         \dot{z}&=z[(\gamma+\beta)xz-\beta x-\min\{\frac{\delta}{z},1\}(cz-c) ] 
\end{aligned}
\right.,
\end{equation}
where $0<\delta<1$. We analyze this system by dividing the phase plane into two regions: $z>\delta$ and $z<\delta$. In the region of $z>\delta$, the equation can be written as 
\begin{equation}
\left\{
\begin{aligned}
         \dot{x}&=x\left[(\gamma+\beta)xz-\gamma z-c\delta \right],  \\
         \dot{z}&=z[(\gamma+\beta)xz-\beta x-c\delta+\frac{c\delta}{z} ]. 
\end{aligned}
\right.
\label{replicator_equation_1}
\end{equation}
Let $e=(x,y,z)$ denote the proportion of three strategies. Equation~(\ref{replicator_equation_1}) has 3 equilibrium solutions: $e_1=(0,0,1)$,
$$e_2=(\frac{2\gamma+\beta-\sqrt{\beta^2-4\delta c(\gamma+\beta)}}{2(\gamma+\beta)},0,\frac{\beta+\sqrt{\beta^2-4\delta c(\gamma+\beta)}}{2(\gamma+\beta)}),$$  $$e_3=(\frac{2\gamma+\beta+\sqrt{\beta^2-4\delta c(\gamma+\beta)}}{2(\gamma+\beta)},0,\frac{\beta-\sqrt{\beta^2-4\delta c(\gamma+\beta)}}{2(\gamma+\beta)}).$$  

In the region of $z<\delta$, equation~(\ref{replicator system}) can be written as
\begin{equation}
\left\{
\begin{aligned}
         \dot{x}&=x\left[(\gamma+\beta)xz-\gamma z-cz \right]  \\
         \dot{z}&=z[(\gamma+\beta)xz-\beta x-cz+c ]
\end{aligned}
\right..
\label{replicator_equation_2}
\end{equation}
This equation has 2 equilibrium solutions: a point $e_4=(\frac{\gamma+c}{\gamma+\beta},0,\frac{\beta-c}{\gamma+\beta})$ and a line $l:z=0$.

Due to the restriction of definition domain, $e_1$, $e_2$, and $e_3$ must be in region of $z>\delta$; $e_4$ and $l$ must be in region of $z<\delta$. $e_1$ and $l$ always exist since they always satisfy the restriction, but the points $e_2$, $e_3$, and $e_4$ do not always exist. Using some basic algebraic manipulations, we can obtain the existing conditions of these three points: 
\\
(1) When $\beta \le c$:
if $\delta \le \frac{\beta^2}{4c(\alpha+\beta)}$, $e_2$ and $e_3$ exist; if $\delta>\frac{\beta^2}{4c(\alpha+\beta)}$, none of $e_2$, $e_3$ and $e_4$ exist.
\\
(2) When $c<\beta<2c$:
if $0<\delta<\frac{\beta-c}{\gamma+\beta}$, only $e_2$ exists; if $\frac{\beta-c}{\gamma+\beta}\le \delta \le \frac{\beta^2}{4c(\alpha+\beta)}$, $e_2$, $e_3$ and $e_4$ all exist; if $\delta > \frac{\beta^2}{4c(\alpha+\beta)}$, only $e_4$ exists.
\\
(3) When $\beta \ge 2c$:
if $0<\delta<\frac{\beta-c}{\gamma+\beta}$, only $e_2$ exists; if $\delta \ge \frac{\beta-c}{\gamma+\beta}$, only $e_4$ exists.

Then, we analyze the stability of these equilibrium solutions. We divide them into two cases, i.e., equation~(\ref{replicator_equation_1}) and equation~(\ref{replicator_equation_2}). 

For equation~(\ref{replicator_equation_1}), let $u=x+z-1$. Equation~(\ref{replicator_equation_1}) can be simplified to
\begin{equation}
\left\{
\begin{aligned}
         \dot{u}&=(\gamma+\beta)(zu^2-z^2u+zu)-\delta cu  \\
         \dot{z}&=-\beta z(u-z+1)-\delta (cz-c)+(\gamma+\beta)z^2(u-z+1)
\end{aligned}
\right.
\label{replicator_u}
\end{equation}
The equilibrium solutions $e_1$, $e_2$ and $e_3$ are transformed to $\tilde{e}_1=(0,1)$, $\tilde{e}_2=(0,\frac{\beta+\sqrt{\beta^2-4\delta c(\gamma+\beta)}}{2(\gamma+\beta)})$ and $\tilde{e}_3=(0,\frac{\beta-\sqrt{\beta^2-4\delta c(\gamma+\beta)}}{2(\gamma+\beta)})$ by $\tilde{e}=(u,z)$. We linearize equation~(\ref{replicator_u}), and get the Jacobian at $u=0$:
\begin{equation}
    J|_{u=0}=\begin{bmatrix}
    -(\gamma+\beta)z^2+(\gamma+\beta)z-\delta c & 0 \\
    * & 2\beta z-\beta-\delta c-3(\gamma+\beta)z^2+2(\gamma+\beta)z
    \end{bmatrix}
    \label{jacobian_1}
\end{equation}
The diagonal entries of $J$ are eigenvalues. Substituting $\tilde{e}_1$, $\tilde{e}_2$ and $\tilde{e}_3$ to equation~(\ref{jacobian_1}), the Jacobian of the system at these fixed points are:
\begin{equation}
\begin{gathered}
    J(\tilde{e}_1)=\begin{bmatrix}
    -\delta c & 0 \\ * & -\delta c-\gamma
    \end{bmatrix},
    J(\tilde{e}_2)=\begin{bmatrix}
   \frac{\gamma\left(\beta+\sqrt{\beta^2-4\delta c(\gamma+\beta)}\right)}{(2\gamma+2\beta)} & 0 \\ * & *
    \end{bmatrix},
    \\ 
    J(\tilde{e}_3)=\begin{bmatrix}
   \frac{\gamma\left(\beta-\sqrt{\beta^2-4\delta c(\gamma+\beta)}\right)}{(2\gamma+2\beta)} & 0 \\ * & *
    \end{bmatrix}.
    \end{gathered}
\end{equation}
Thus, $J(\tilde{e}_1)$ has two negative eigenvalues; $J(\tilde{e}_2)$ and $J(\tilde{e}_3)$ both have at least one positive eigenvalue. So, $e_1$ is stable but $e_2$ and $e_3$ are unstable.

For equation~(\ref{replicator_equation_2}), the Jacobian of this system is
\begin{equation}
    J=\begin{bmatrix}
    2(\gamma+\beta)xz-\gamma z -cz & (\gamma+\beta)x^2-\gamma x-cx \\
    (\gamma+\beta)z^2-\beta z & 2(\gamma+\beta)xz-\beta x-2cz+c
    \end{bmatrix}
    \label{jacobian_2}
\end{equation}
Substituting $e_4$ and $l$ to equation~(\ref{jacobian_2}), The Jacobian of the system at these solutions are:
\begin{equation}
    J(e_4)=\begin{bmatrix}
    \frac{(\gamma+c)(\beta-c)}{\gamma+\beta} & 0 \\ * & \frac{\gamma(\beta-c)}{\gamma+\beta}
    \end{bmatrix},
    J(l)=\begin{bmatrix}
    0  & (\gamma+\beta)x^2-\gamma x-cx \\ 0 & -\beta x+c
    \end{bmatrix}
\end{equation}
Since $\beta>c$ when $e_4$ exists, $J(e_4)$ has two positive eigenvalues. Thus, $e_4$ is unstable. Since each point on $l$ is a fixed point, we only need to check the stability in the direction of $z$, i.e., the sign of $-\beta x+c$. If a point on $l$ satisfies $x>c/\beta$, this point is stable. Thus, the stable manifold of $l$ is $\{e=(x,1-x,0)|c/\beta<x\le 1\}$.
    
\section*{Acknowledgments}
G.W. and L.W. gratefully acknowledge the support from the National Natural Science Foundation of China (NSFC 62036002) and PKU-Baidu Fund (2020BD017). Q.S. acknowledges support by the Simons Foundation Math+X Grant to the University of Pennsylvania.

\bibliography{reference}

\begin{thebibliography}{50}%
\makeatletter
\providecommand \@ifxundefined [1]{%
 \@ifx{#1\undefined}
}%
\providecommand \@ifnum [1]{%
 \ifnum #1\expandafter \@firstoftwo
 \else \expandafter \@secondoftwo
 \fi
}%
\providecommand \@ifx [1]{%
 \ifx #1\expandafter \@firstoftwo
 \else \expandafter \@secondoftwo
 \fi
}%
\providecommand \natexlab [1]{#1}%
\providecommand \enquote  [1]{``#1''}%
\providecommand \bibnamefont  [1]{#1}%
\providecommand \bibfnamefont [1]{#1}%
\providecommand \citenamefont [1]{#1}%
\providecommand \href@noop [0]{\@secondoftwo}%
\providecommand \href [0]{\begingroup \@sanitize@url \@href}%
\providecommand \@href[1]{\@@startlink{#1}\@@href}%
\providecommand \@@href[1]{\endgroup#1\@@endlink}%
\providecommand \@sanitize@url [0]{\catcode `\\12\catcode `\$12\catcode
  `\&12\catcode `\#12\catcode `\^12\catcode `\_12\catcode `\%12\relax}%
\providecommand \@@startlink[1]{}%
\providecommand \@@endlink[0]{}%
\providecommand \url  [0]{\begingroup\@sanitize@url \@url }%
\providecommand \@url [1]{\endgroup\@href {#1}{\urlprefix }}%
\providecommand \urlprefix  [0]{URL }%
\providecommand \Eprint [0]{\href }%
\providecommand \doibase [0]{https://doi.org/}%
\providecommand \selectlanguage [0]{\@gobble}%
\providecommand \bibinfo  [0]{\@secondoftwo}%
\providecommand \bibfield  [0]{\@secondoftwo}%
\providecommand \translation [1]{[#1]}%
\providecommand \BibitemOpen [0]{}%
\providecommand \bibitemStop [0]{}%
\providecommand \bibitemNoStop [0]{.\EOS\space}%
\providecommand \EOS [0]{\spacefactor3000\relax}%
\providecommand \BibitemShut  [1]{\csname bibitem#1\endcsname}%
\let\auto@bib@innerbib\@empty
\bibitem [{\citenamefont {Tyler}(2006)}]{Tyler2006}%
  \BibitemOpen
  \bibfield  {author} {\bibinfo {author} {\bibfnamefont {T.~R.}\ \bibnamefont
  {Tyler}},\ }\href@noop {} {\emph {\bibinfo {title} {{Why people obey the
  law}}}}\ (\bibinfo  {publisher} {Princeton University Press},\ \bibinfo
  {address} {Princeton},\ \bibinfo {year} {2006})\BibitemShut {NoStop}%
\bibitem [{\citenamefont {Dan-Cohen}(1992)}]{Dan-Cohen1992}%
  \BibitemOpen
  \bibfield  {author} {\bibinfo {author} {\bibfnamefont {M.}~\bibnamefont
  {Dan-Cohen}},\ }\bibfield  {title} {\bibinfo {title} {{Responsibility and the
  boundaries of the self}},\ }\href {https://doi.org/10.2307/1341517}
  {\bibfield  {journal} {\bibinfo  {journal} {Harv. Law Rev.}\ }\textbf
  {\bibinfo {volume} {105}},\ \bibinfo {pages} {959} (\bibinfo {year}
  {1992})}\BibitemShut {NoStop}%
\bibitem [{\citenamefont {Kornhauser}\ and\ \citenamefont
  {Revesz}(1994)}]{Kornhauser1994}%
  \BibitemOpen
  \bibfield  {author} {\bibinfo {author} {\bibfnamefont {L.~A.}\ \bibnamefont
  {Kornhauser}}\ and\ \bibinfo {author} {\bibfnamefont {R.~L.}\ \bibnamefont
  {Revesz}},\ }\bibfield  {title} {\bibinfo {title} {{Multidefendant
  settlements: The impact of joint and several liability}},\ }\href
  {https://doi.org/10.1086/467916} {\bibfield  {journal} {\bibinfo  {journal}
  {J. Leg. Stud.}\ }\textbf {\bibinfo {volume} {23}},\ \bibinfo {pages} {41}
  (\bibinfo {year} {1994})}\BibitemShut {NoStop}%
\bibitem [{\citenamefont {Ren}(1997)}]{Ren1997}%
  \BibitemOpen
  \bibfield  {author} {\bibinfo {author} {\bibfnamefont {X.}~\bibnamefont
  {Ren}},\ }\href@noop {} {\emph {\bibinfo {title} {{Tradition of the law and
  law of the tradition: law, state, and social control in China}}}}\ (\bibinfo
  {publisher} {Greenwood Press},\ \bibinfo {address} {London},\ \bibinfo {year}
  {1997})\BibitemShut {NoStop}%
\bibitem [{\citenamefont {Alchian}\ and\ \citenamefont
  {Demsetz}(1972)}]{Alchian1972}%
  \BibitemOpen
  \bibfield  {author} {\bibinfo {author} {\bibfnamefont {A.~A.}\ \bibnamefont
  {Alchian}}\ and\ \bibinfo {author} {\bibfnamefont {H.}~\bibnamefont
  {Demsetz}},\ }\bibfield  {title} {\bibinfo {title} {{Production, information
  costs, and economic organization}},\ }\href
  {http://www.jstor.org/stable/1815199} {\bibfield  {journal} {\bibinfo
  {journal} {Am. Econ. Rev.}\ }\textbf {\bibinfo {volume} {62}},\ \bibinfo
  {pages} {777} (\bibinfo {year} {1972})}\BibitemShut {NoStop}%
\bibitem [{\citenamefont {Clutton-Brock}\ and\ \citenamefont
  {Parker}(1995)}]{Clutton-Brock1995}%
  \BibitemOpen
  \bibfield  {author} {\bibinfo {author} {\bibfnamefont {T.~H.}\ \bibnamefont
  {Clutton-Brock}}\ and\ \bibinfo {author} {\bibfnamefont {G.~A.}\ \bibnamefont
  {Parker}},\ }\bibfield  {title} {\bibinfo {title} {{Punishment in animal
  society}},\ }\href {https://doi.org/10.1038/373209a0} {\bibfield  {journal}
  {\bibinfo  {journal} {Nature}\ }\textbf {\bibinfo {volume} {373}},\ \bibinfo
  {pages} {209} (\bibinfo {year} {1995})}\BibitemShut {NoStop}%
\bibitem [{\citenamefont {Allport}(1954)}]{Allport1954}%
  \BibitemOpen
  \bibfield  {author} {\bibinfo {author} {\bibfnamefont {G.~W.}\ \bibnamefont
  {Allport}},\ }\href@noop {} {\emph {\bibinfo {title} {{The Nature of
  Prejudice}}}}\ (\bibinfo  {publisher} {Addison-Wesley},\ \bibinfo {address}
  {Oxford},\ \bibinfo {year} {1954})\BibitemShut {NoStop}%
\bibitem [{\citenamefont {Akerlof}(1970)}]{Akerlof1970}%
  \BibitemOpen
  \bibfield  {author} {\bibinfo {author} {\bibfnamefont {G.~A.}\ \bibnamefont
  {Akerlof}},\ }\bibfield  {title} {\bibinfo {title} {{The market for
  ``lemons": Quality uncertainty and the market mechanism}},\ }\href
  {https://doi.org/10.1016/b978-0-12-214850-7.50022-x} {\bibfield  {journal}
  {\bibinfo  {journal} {Q. J. Econ.}\ }\textbf {\bibinfo {volume} {84}},\
  \bibinfo {pages} {488} (\bibinfo {year} {1970})}\BibitemShut {NoStop}%
\bibitem [{\citenamefont {Hossain}(1988)}]{Hossain1988}%
  \BibitemOpen
  \bibfield  {author} {\bibinfo {author} {\bibfnamefont {M.}~\bibnamefont
  {Hossain}},\ }\href@noop {} {\emph {\bibinfo {title} {{Credit for alleviation
  of rural poverty: The Grameen Bank in Bangladesh}}}}\ (\bibinfo  {publisher}
  {Intl. Food Policy Res. Inst.},\ \bibinfo {year} {1988})\BibitemShut
  {NoStop}%
\bibitem [{\citenamefont {Ghatak}\ and\ \citenamefont
  {Guinnane}(1999)}]{Ghatak1999}%
  \BibitemOpen
  \bibfield  {author} {\bibinfo {author} {\bibfnamefont {M.}~\bibnamefont
  {Ghatak}}\ and\ \bibinfo {author} {\bibfnamefont {T.~W.}\ \bibnamefont
  {Guinnane}},\ }\bibfield  {title} {\bibinfo {title} {{The economics of
  lending with joint liability: Theory and practice}},\ }\href
  {https://doi.org/10.1016/S0304-3878(99)00041-3} {\bibfield  {journal}
  {\bibinfo  {journal} {J. Dev. Econ.}\ }\textbf {\bibinfo {volume} {60}},\
  \bibinfo {pages} {195} (\bibinfo {year} {1999})}\BibitemShut {NoStop}%
\bibitem [{\citenamefont {Hilbe}\ \emph {et~al.}(2017)\citenamefont {Hilbe},
  \citenamefont {Martinez-Vaquero}, \citenamefont {Chatterjee},\ and\
  \citenamefont {Nowak}}]{Hilbe2017}%
  \BibitemOpen
  \bibfield  {author} {\bibinfo {author} {\bibfnamefont {C.}~\bibnamefont
  {Hilbe}}, \bibinfo {author} {\bibfnamefont {L.~A.}\ \bibnamefont
  {Martinez-Vaquero}}, \bibinfo {author} {\bibfnamefont {K.}~\bibnamefont
  {Chatterjee}},\ and\ \bibinfo {author} {\bibfnamefont {M.~A.}\ \bibnamefont
  {Nowak}},\ }\bibfield  {title} {\bibinfo {title} {{Memory-n strategies of
  direct reciprocity}},\ }\href {https://doi.org/10.1073/pnas.1621239114}
  {\bibfield  {journal} {\bibinfo  {journal} {Proc. Natl. Acad. Sci. U.S.A.}\
  }\textbf {\bibinfo {volume} {114}},\ \bibinfo {pages} {4715} (\bibinfo {year}
  {2017})}\BibitemShut {NoStop}%
\bibitem [{\citenamefont {Nowak}\ and\ \citenamefont
  {Sigmund}(1993)}]{Nowak1993}%
  \BibitemOpen
  \bibfield  {author} {\bibinfo {author} {\bibfnamefont {M.~A.}\ \bibnamefont
  {Nowak}}\ and\ \bibinfo {author} {\bibfnamefont {K.}~\bibnamefont
  {Sigmund}},\ }\bibfield  {title} {\bibinfo {title} {{A strategy of win-stay,
  lose-shift that outperforms tit-for-tat in the Prisoner's Dilemma game}},\
  }\href {https://doi.org/10.1038/364056a0} {\bibfield  {journal} {\bibinfo
  {journal} {Nature}\ }\textbf {\bibinfo {volume} {364}},\ \bibinfo {pages}
  {56} (\bibinfo {year} {1993})}\BibitemShut {NoStop}%
\bibitem [{\citenamefont {Nowak}\ and\ \citenamefont
  {Sigmund}(2005)}]{Nowak2005}%
  \BibitemOpen
  \bibfield  {author} {\bibinfo {author} {\bibfnamefont {M.~A.}\ \bibnamefont
  {Nowak}}\ and\ \bibinfo {author} {\bibfnamefont {K.}~\bibnamefont
  {Sigmund}},\ }\bibfield  {title} {\bibinfo {title} {{Evolution of indirect
  reciprocity}},\ }\href {https://doi.org/10.1038/nature04131} {\bibfield
  {journal} {\bibinfo  {journal} {Nature}\ }\textbf {\bibinfo {volume} {437}},\
  \bibinfo {pages} {1291} (\bibinfo {year} {2005})}\BibitemShut {NoStop}%
\bibitem [{\citenamefont {Hilbe}\ \emph {et~al.}(2018)\citenamefont {Hilbe},
  \citenamefont {Schmid}, \citenamefont {Tkadlec}, \citenamefont {Chatterjee},\
  and\ \citenamefont {Nowak}}]{Hilbe2018}%
  \BibitemOpen
  \bibfield  {author} {\bibinfo {author} {\bibfnamefont {C.}~\bibnamefont
  {Hilbe}}, \bibinfo {author} {\bibfnamefont {L.}~\bibnamefont {Schmid}},
  \bibinfo {author} {\bibfnamefont {J.}~\bibnamefont {Tkadlec}}, \bibinfo
  {author} {\bibfnamefont {K.}~\bibnamefont {Chatterjee}},\ and\ \bibinfo
  {author} {\bibfnamefont {M.~A.}\ \bibnamefont {Nowak}},\ }\bibfield  {title}
  {\bibinfo {title} {{Indirect reciprocity with private, noisy, and incomplete
  information}},\ }\href {https://doi.org/10.1073/pnas.1810565115} {\bibfield
  {journal} {\bibinfo  {journal} {Proc. Natl. Acad. Sci. U.S.A.}\ }\textbf
  {\bibinfo {volume} {115}},\ \bibinfo {pages} {12241} (\bibinfo {year}
  {2018})}\BibitemShut {NoStop}%
\bibitem [{\citenamefont {Nowak}\ and\ \citenamefont
  {Sigmund}(1998)}]{Nowak1998}%
  \BibitemOpen
  \bibfield  {author} {\bibinfo {author} {\bibfnamefont {M.~A.}\ \bibnamefont
  {Nowak}}\ and\ \bibinfo {author} {\bibfnamefont {K.}~\bibnamefont
  {Sigmund}},\ }\bibfield  {title} {\bibinfo {title} {{Evolution of indirect
  reciprocity by image scoring}},\ }\href {https://doi.org/10.1038/31225}
  {\bibfield  {journal} {\bibinfo  {journal} {Nature}\ }\textbf {\bibinfo
  {volume} {393}},\ \bibinfo {pages} {573} (\bibinfo {year}
  {1998})}\BibitemShut {NoStop}%
\bibitem [{\citenamefont {Santos}\ \emph {et~al.}(2018)\citenamefont {Santos},
  \citenamefont {Santos},\ and\ \citenamefont {Pacheco}}]{Santos2018}%
  \BibitemOpen
  \bibfield  {author} {\bibinfo {author} {\bibfnamefont {F.~P.}\ \bibnamefont
  {Santos}}, \bibinfo {author} {\bibfnamefont {F.~C.}\ \bibnamefont {Santos}},\
  and\ \bibinfo {author} {\bibfnamefont {J.~M.}\ \bibnamefont {Pacheco}},\
  }\bibfield  {title} {\bibinfo {title} {{Social norm complexity and past
  reputations in the evolution of cooperation}},\ }\href
  {https://doi.org/10.1038/nature25763} {\bibfield  {journal} {\bibinfo
  {journal} {Nature}\ }\textbf {\bibinfo {volume} {555}},\ \bibinfo {pages}
  {242} (\bibinfo {year} {2018})}\BibitemShut {NoStop}%
\bibitem [{\citenamefont {Ohtsuki}\ \emph {et~al.}(2006)\citenamefont
  {Ohtsuki}, \citenamefont {Hauert}, \citenamefont {Lieberman},\ and\
  \citenamefont {Nowak}}]{Ohtsuki2006}%
  \BibitemOpen
  \bibfield  {author} {\bibinfo {author} {\bibfnamefont {H.}~\bibnamefont
  {Ohtsuki}}, \bibinfo {author} {\bibfnamefont {C.}~\bibnamefont {Hauert}},
  \bibinfo {author} {\bibfnamefont {E.}~\bibnamefont {Lieberman}},\ and\
  \bibinfo {author} {\bibfnamefont {M.~A.}\ \bibnamefont {Nowak}},\ }\bibfield
  {title} {\bibinfo {title} {{A simple rule for the evolution of cooperation on
  graphs and social networks}},\ }\href {https://doi.org/10.1038/nature04605}
  {\bibfield  {journal} {\bibinfo  {journal} {Nature}\ }\textbf {\bibinfo
  {volume} {441}},\ \bibinfo {pages} {502} (\bibinfo {year}
  {2006})}\BibitemShut {NoStop}%
\bibitem [{\citenamefont {Su}\ \emph {et~al.}(2019{\natexlab{a}})\citenamefont
  {Su}, \citenamefont {McAvoy}, \citenamefont {Wang},\ and\ \citenamefont
  {Nowak}}]{Su2019a}%
  \BibitemOpen
  \bibfield  {author} {\bibinfo {author} {\bibfnamefont {Q.}~\bibnamefont
  {Su}}, \bibinfo {author} {\bibfnamefont {A.}~\bibnamefont {McAvoy}}, \bibinfo
  {author} {\bibfnamefont {L.}~\bibnamefont {Wang}},\ and\ \bibinfo {author}
  {\bibfnamefont {M.~A.}\ \bibnamefont {Nowak}},\ }\bibfield  {title} {\bibinfo
  {title} {{Evolutionary dynamics with game transitions}},\ }\href
  {https://doi.org/10.1073/pnas.1908936116} {\bibfield  {journal} {\bibinfo
  {journal} {Proc. Natl. Acad. Sci. U.S.A.}\ }\textbf {\bibinfo {volume}
  {116}},\ \bibinfo {pages} {25398} (\bibinfo {year}
  {2019}{\natexlab{a}})}\BibitemShut {NoStop}%
\bibitem [{\citenamefont {Nowak}\ and\ \citenamefont {May}(1992)}]{Nowak1992}%
  \BibitemOpen
  \bibfield  {author} {\bibinfo {author} {\bibfnamefont {M.~A.}\ \bibnamefont
  {Nowak}}\ and\ \bibinfo {author} {\bibfnamefont {R.~M.}\ \bibnamefont
  {May}},\ }\bibfield  {title} {\bibinfo {title} {{Evolutionary games and
  spatial chaos}},\ }\href {https://doi.org/10.1038/359826a0} {\bibfield
  {journal} {\bibinfo  {journal} {Nature}\ }\textbf {\bibinfo {volume} {359}},\
  \bibinfo {pages} {826} (\bibinfo {year} {1992})}\BibitemShut {NoStop}%
\bibitem [{\citenamefont {Allen}\ \emph {et~al.}(2017)\citenamefont {Allen},
  \citenamefont {Lippner}, \citenamefont {Chen}, \citenamefont {Fotouhi},
  \citenamefont {Momeni}, \citenamefont {Yau},\ and\ \citenamefont
  {Nowak}}]{Allen2017}%
  \BibitemOpen
  \bibfield  {author} {\bibinfo {author} {\bibfnamefont {B.}~\bibnamefont
  {Allen}}, \bibinfo {author} {\bibfnamefont {G.}~\bibnamefont {Lippner}},
  \bibinfo {author} {\bibfnamefont {Y.~T.}\ \bibnamefont {Chen}}, \bibinfo
  {author} {\bibfnamefont {B.}~\bibnamefont {Fotouhi}}, \bibinfo {author}
  {\bibfnamefont {N.}~\bibnamefont {Momeni}}, \bibinfo {author} {\bibfnamefont
  {S.~T.}\ \bibnamefont {Yau}},\ and\ \bibinfo {author} {\bibfnamefont {M.~A.}\
  \bibnamefont {Nowak}},\ }\bibfield  {title} {\bibinfo {title} {{Evolutionary
  dynamics on any population structure}},\ }\href
  {https://doi.org/10.1038/nature21723} {\bibfield  {journal} {\bibinfo
  {journal} {Nature}\ }\textbf {\bibinfo {volume} {544}},\ \bibinfo {pages}
  {227} (\bibinfo {year} {2017})}\BibitemShut {NoStop}%
\bibitem [{\citenamefont {McAvoy}\ \emph {et~al.}(2020)\citenamefont {McAvoy},
  \citenamefont {Allen},\ and\ \citenamefont {Nowak}}]{McAvoy2020}%
  \BibitemOpen
  \bibfield  {author} {\bibinfo {author} {\bibfnamefont {A.}~\bibnamefont
  {McAvoy}}, \bibinfo {author} {\bibfnamefont {B.}~\bibnamefont {Allen}},\ and\
  \bibinfo {author} {\bibfnamefont {M.~A.}\ \bibnamefont {Nowak}},\ }\bibfield
  {title} {\bibinfo {title} {{Social goods dilemmas in heterogeneous
  societies}},\ }\href {https://doi.org/10.1038/s41562-020-0881-2} {\bibfield
  {journal} {\bibinfo  {journal} {Nat. Hum. Behav.}\ }\textbf {\bibinfo
  {volume} {4}},\ \bibinfo {pages} {819} (\bibinfo {year} {2020})}\BibitemShut
  {NoStop}%
\bibitem [{\citenamefont {Hauert}\ \emph {et~al.}(2007)\citenamefont {Hauert},
  \citenamefont {Traulsen}, \citenamefont {Brandt}, \citenamefont {Nowak},\
  and\ \citenamefont {Sigmund}}]{Hauert2007}%
  \BibitemOpen
  \bibfield  {author} {\bibinfo {author} {\bibfnamefont {C.}~\bibnamefont
  {Hauert}}, \bibinfo {author} {\bibfnamefont {A.}~\bibnamefont {Traulsen}},
  \bibinfo {author} {\bibfnamefont {H.}~\bibnamefont {Brandt}}, \bibinfo
  {author} {\bibfnamefont {M.~A.}\ \bibnamefont {Nowak}},\ and\ \bibinfo
  {author} {\bibfnamefont {K.}~\bibnamefont {Sigmund}},\ }\bibfield  {title}
  {\bibinfo {title} {{Via freedom to coercion: The emergence of costly
  punishment}},\ }\href {https://doi.org/10.1126/science.1141588} {\bibfield
  {journal} {\bibinfo  {journal} {Science}\ }\textbf {\bibinfo {volume}
  {316}},\ \bibinfo {pages} {1905} (\bibinfo {year} {2007})}\BibitemShut
  {NoStop}%
\bibitem [{\citenamefont {Helbing}\ \emph {et~al.}(2010)\citenamefont
  {Helbing}, \citenamefont {Szolnoki}, \citenamefont {Perc},\ and\
  \citenamefont {Szab{\'{o}}}}]{Helbing2010}%
  \BibitemOpen
  \bibfield  {author} {\bibinfo {author} {\bibfnamefont {D.}~\bibnamefont
  {Helbing}}, \bibinfo {author} {\bibfnamefont {A.}~\bibnamefont {Szolnoki}},
  \bibinfo {author} {\bibfnamefont {M.}~\bibnamefont {Perc}},\ and\ \bibinfo
  {author} {\bibfnamefont {G.}~\bibnamefont {Szab{\'{o}}}},\ }\bibfield
  {title} {\bibinfo {title} {{Punish, but not too hard: How costly punishment
  spreads in the spatial public goods game}},\ }\href
  {https://doi.org/10.1088/1367-2630/12/8/083005} {\bibfield  {journal}
  {\bibinfo  {journal} {New J. Phys.}\ }\textbf {\bibinfo {volume} {12}},\
  \bibinfo {pages} {083005} (\bibinfo {year} {2010})}\BibitemShut {NoStop}%
\bibitem [{\citenamefont {Szolnoki}\ and\ \citenamefont
  {Perc}(2017)}]{Szolnoki2017}%
  \BibitemOpen
  \bibfield  {author} {\bibinfo {author} {\bibfnamefont {A.}~\bibnamefont
  {Szolnoki}}\ and\ \bibinfo {author} {\bibfnamefont {M.}~\bibnamefont
  {Perc}},\ }\bibfield  {title} {\bibinfo {title} {{Second-order free-riding on
  antisocial punishment restores the effectiveness of prosocial punishment}},\
  }\href {https://doi.org/10.1103/PhysRevX.7.041027} {\bibfield  {journal}
  {\bibinfo  {journal} {Phys. Rev. X}\ }\textbf {\bibinfo {volume} {7}},\
  \bibinfo {pages} {041027} (\bibinfo {year} {2017})}\BibitemShut {NoStop}%
\bibitem [{\citenamefont {Henrich}\ \emph {et~al.}(2006)\citenamefont
  {Henrich}, \citenamefont {McElreath}, \citenamefont {Barr}, \citenamefont
  {Ensminger}, \citenamefont {Barrett}, \citenamefont {Bolyanatz},
  \citenamefont {Cardaroas}, \citenamefont {Gurven}, \citenamefont {Gwako},
  \citenamefont {Henrich}, \citenamefont {Lesoronol}, \citenamefont {Marlowe},
  \citenamefont {Tracer},\ and\ \citenamefont {Ziker}}]{Henrich2006}%
  \BibitemOpen
  \bibfield  {author} {\bibinfo {author} {\bibfnamefont {J.}~\bibnamefont
  {Henrich}}, \bibinfo {author} {\bibfnamefont {R.}~\bibnamefont {McElreath}},
  \bibinfo {author} {\bibfnamefont {A.}~\bibnamefont {Barr}}, \bibinfo {author}
  {\bibfnamefont {J.}~\bibnamefont {Ensminger}}, \bibinfo {author}
  {\bibfnamefont {C.}~\bibnamefont {Barrett}}, \bibinfo {author} {\bibfnamefont
  {A.}~\bibnamefont {Bolyanatz}}, \bibinfo {author} {\bibfnamefont {J.~C.}\
  \bibnamefont {Cardaroas}}, \bibinfo {author} {\bibfnamefont {M.}~\bibnamefont
  {Gurven}}, \bibinfo {author} {\bibfnamefont {E.}~\bibnamefont {Gwako}},
  \bibinfo {author} {\bibfnamefont {N.}~\bibnamefont {Henrich}}, \bibinfo
  {author} {\bibfnamefont {C.}~\bibnamefont {Lesoronol}}, \bibinfo {author}
  {\bibfnamefont {F.}~\bibnamefont {Marlowe}}, \bibinfo {author} {\bibfnamefont
  {D.}~\bibnamefont {Tracer}},\ and\ \bibinfo {author} {\bibfnamefont
  {J.}~\bibnamefont {Ziker}},\ }\bibfield  {title} {\bibinfo {title} {{Costly
  punishment across human societies}},\ }\href
  {https://doi.org/10.1126/science.1127333} {\bibfield  {journal} {\bibinfo
  {journal} {Science}\ }\textbf {\bibinfo {volume} {312}},\ \bibinfo {pages}
  {1767} (\bibinfo {year} {2006})}\BibitemShut {NoStop}%
\bibitem [{\citenamefont {Ohtsuki}\ \emph {et~al.}(2009)\citenamefont
  {Ohtsuki}, \citenamefont {Iwasa},\ and\ \citenamefont {Nowak}}]{Ohtsuki2009}%
  \BibitemOpen
  \bibfield  {author} {\bibinfo {author} {\bibfnamefont {H.}~\bibnamefont
  {Ohtsuki}}, \bibinfo {author} {\bibfnamefont {Y.}~\bibnamefont {Iwasa}},\
  and\ \bibinfo {author} {\bibfnamefont {M.~A.}\ \bibnamefont {Nowak}},\
  }\bibfield  {title} {\bibinfo {title} {{Indirect reciprocity provides only a
  narrow margin of efficiency for costly punishment}},\ }\href
  {https://doi.org/10.1038/nature07601} {\bibfield  {journal} {\bibinfo
  {journal} {Nature}\ }\textbf {\bibinfo {volume} {457}},\ \bibinfo {pages}
  {79} (\bibinfo {year} {2009})}\BibitemShut {NoStop}%
\bibitem [{\citenamefont {Colman}(2006)}]{Colman2006}%
  \BibitemOpen
  \bibfield  {author} {\bibinfo {author} {\bibfnamefont {A.~M.}\ \bibnamefont
  {Colman}},\ }\bibfield  {title} {\bibinfo {title} {{The puzzle of
  cooperation}},\ }\href {https://doi.org/10.1038/440744b} {\bibfield
  {journal} {\bibinfo  {journal} {Nature}\ }\textbf {\bibinfo {volume} {440}},\
  \bibinfo {pages} {744} (\bibinfo {year} {2006})}\BibitemShut {NoStop}%
\bibitem [{\citenamefont {Nowak}(2006{\natexlab{a}})}]{Nowak2006book}%
  \BibitemOpen
  \bibfield  {author} {\bibinfo {author} {\bibfnamefont {M.~A.}\ \bibnamefont
  {Nowak}},\ }\href@noop {} {\emph {\bibinfo {title} {{Evolutionary dynamics:
  exploring the equations of life}}}}\ (\bibinfo  {publisher} {Harvard
  University Press},\ \bibinfo {address} {Cambridge},\ \bibinfo {year}
  {2006})\BibitemShut {NoStop}%
\bibitem [{\citenamefont {Wu}\ \emph {et~al.}(2013)\citenamefont {Wu},
  \citenamefont {Garc{\'{i}}a}, \citenamefont {Hauert},\ and\ \citenamefont
  {Traulsen}}]{Wu2013}%
  \BibitemOpen
  \bibfield  {author} {\bibinfo {author} {\bibfnamefont {B.}~\bibnamefont
  {Wu}}, \bibinfo {author} {\bibfnamefont {J.}~\bibnamefont {Garc{\'{i}}a}},
  \bibinfo {author} {\bibfnamefont {C.}~\bibnamefont {Hauert}},\ and\ \bibinfo
  {author} {\bibfnamefont {A.}~\bibnamefont {Traulsen}},\ }\bibfield  {title}
  {\bibinfo {title} {{Extrapolating weak selection in evolutionary games}},\
  }\href {https://doi.org/10.1371/journal.pcbi.1003381} {\bibfield  {journal}
  {\bibinfo  {journal} {PLoS Comput. Biol.}\ }\textbf {\bibinfo {volume} {9}},\
  \bibinfo {pages} {e1003381} (\bibinfo {year} {2013})}\BibitemShut {NoStop}%
\bibitem [{\citenamefont {Nowak}(2006{\natexlab{b}})}]{Nowak2006}%
  \BibitemOpen
  \bibfield  {author} {\bibinfo {author} {\bibfnamefont {M.~A.}\ \bibnamefont
  {Nowak}},\ }\bibfield  {title} {\bibinfo {title} {{Five rules for the
  evolution of cooperation}},\ }\href {https://doi.org/10.1126/science.1133755}
  {\bibfield  {journal} {\bibinfo  {journal} {Science}\ }\textbf {\bibinfo
  {volume} {314}},\ \bibinfo {pages} {1560} (\bibinfo {year}
  {2006}{\natexlab{b}})}\BibitemShut {NoStop}%
\bibitem [{\citenamefont {Nowak}\ \emph {et~al.}(2004)\citenamefont {Nowak},
  \citenamefont {Sasaki}, \citenamefont {Taylor},\ and\ \citenamefont
  {Fudenherg}}]{Nowak2004}%
  \BibitemOpen
  \bibfield  {author} {\bibinfo {author} {\bibfnamefont {M.~A.}\ \bibnamefont
  {Nowak}}, \bibinfo {author} {\bibfnamefont {A.}~\bibnamefont {Sasaki}},
  \bibinfo {author} {\bibfnamefont {C.}~\bibnamefont {Taylor}},\ and\ \bibinfo
  {author} {\bibfnamefont {D.}~\bibnamefont {Fudenherg}},\ }\bibfield  {title}
  {\bibinfo {title} {{Emergence of cooperation and evolutionary stability in
  finite populations}},\ }\href {https://doi.org/10.1038/nature02414}
  {\bibfield  {journal} {\bibinfo  {journal} {Nature}\ }\textbf {\bibinfo
  {volume} {428}},\ \bibinfo {pages} {646} (\bibinfo {year}
  {2004})}\BibitemShut {NoStop}%
\bibitem [{\citenamefont {Fudenberg}\ and\ \citenamefont
  {Imhof}(2006)}]{Fudenberg2006}%
  \BibitemOpen
  \bibfield  {author} {\bibinfo {author} {\bibfnamefont {D.}~\bibnamefont
  {Fudenberg}}\ and\ \bibinfo {author} {\bibfnamefont {L.~A.}\ \bibnamefont
  {Imhof}},\ }\bibfield  {title} {\bibinfo {title} {{Imitation processes with
  small mutations}},\ }\href {https://doi.org/10.1016/j.jet.2005.04.006}
  {\bibfield  {journal} {\bibinfo  {journal} {J. Econ. Theory}\ }\textbf
  {\bibinfo {volume} {131}},\ \bibinfo {pages} {251} (\bibinfo {year}
  {2006})}\BibitemShut {NoStop}%
\bibitem [{\citenamefont {Wu}\ \emph {et~al.}(2015)\citenamefont {Wu},
  \citenamefont {Bauer}, \citenamefont {Galla},\ and\ \citenamefont
  {Traulsen}}]{Wu2015}%
  \BibitemOpen
  \bibfield  {author} {\bibinfo {author} {\bibfnamefont {B.}~\bibnamefont
  {Wu}}, \bibinfo {author} {\bibfnamefont {B.}~\bibnamefont {Bauer}}, \bibinfo
  {author} {\bibfnamefont {T.}~\bibnamefont {Galla}},\ and\ \bibinfo {author}
  {\bibfnamefont {A.}~\bibnamefont {Traulsen}},\ }\bibfield  {title} {\bibinfo
  {title} {{Fitness-based models and pairwise comparison models of evolutionary
  games are typically different - Even in unstructured populations}},\ }\href
  {https://doi.org/10.1088/1367-2630/17/2/023043} {\bibfield  {journal}
  {\bibinfo  {journal} {New J. Phys.}\ }\textbf {\bibinfo {volume} {17}},\
  \bibinfo {pages} {023043} (\bibinfo {year} {2015})}\BibitemShut {NoStop}%
\bibitem [{\citenamefont {Traulsen}\ \emph {et~al.}(2005)\citenamefont
  {Traulsen}, \citenamefont {Claussen},\ and\ \citenamefont
  {Hauert}}]{Traulsen2005}%
  \BibitemOpen
  \bibfield  {author} {\bibinfo {author} {\bibfnamefont {A.}~\bibnamefont
  {Traulsen}}, \bibinfo {author} {\bibfnamefont {J.~C.}\ \bibnamefont
  {Claussen}},\ and\ \bibinfo {author} {\bibfnamefont {C.}~\bibnamefont
  {Hauert}},\ }\bibfield  {title} {\bibinfo {title} {{Coevolutionary dynamics:
  From finite to infinite populations}},\ }\href
  {https://doi.org/10.1103/PhysRevLett.95.238701} {\bibfield  {journal}
  {\bibinfo  {journal} {Phys. Rev. Lett.}\ }\textbf {\bibinfo {volume} {95}},\
  \bibinfo {pages} {238701} (\bibinfo {year} {2005})}\BibitemShut {NoStop}%
\bibitem [{\citenamefont {Molina}\ and\ \citenamefont
  {Earn}(2021)}]{Molina2021}%
  \BibitemOpen
  \bibfield  {author} {\bibinfo {author} {\bibfnamefont {C.}~\bibnamefont
  {Molina}}\ and\ \bibinfo {author} {\bibfnamefont {D.~J.~D.}\ \bibnamefont
  {Earn}},\ }\bibfield  {title} {\bibinfo {title} {{On inferring evolutionary
  stability in finite populations using infinite population models}},\ }\href
  {https://doi.org/10.1007/s00285-021-01636-9} {\bibfield  {journal} {\bibinfo
  {journal} {J. Math. Biol.}\ }\textbf {\bibinfo {volume} {83}},\ \bibinfo
  {pages} {21} (\bibinfo {year} {2021})}\BibitemShut {NoStop}%
\bibitem [{\citenamefont {Traulsen}\ and\ \citenamefont
  {Nowak}(2006)}]{Traulsen2006}%
  \BibitemOpen
  \bibfield  {author} {\bibinfo {author} {\bibfnamefont {A.}~\bibnamefont
  {Traulsen}}\ and\ \bibinfo {author} {\bibfnamefont {M.~A.}\ \bibnamefont
  {Nowak}},\ }\bibfield  {title} {\bibinfo {title} {{Evolution of cooperation
  by multilevel selection}},\ }\href {https://doi.org/10.1073/pnas.0602530103}
  {\bibfield  {journal} {\bibinfo  {journal} {Proc. Natl. Acad. Sci. U.S.A.}\
  }\textbf {\bibinfo {volume} {103}},\ \bibinfo {pages} {10952} (\bibinfo
  {year} {2006})}\BibitemShut {NoStop}%
\bibitem [{\citenamefont {Antal}\ \emph {et~al.}(2009)\citenamefont {Antal},
  \citenamefont {Ohtsuki}, \citenamefont {Wakeley}, \citenamefont {Taylor},\
  and\ \citenamefont {Nowak}}]{Antal2009}%
  \BibitemOpen
  \bibfield  {author} {\bibinfo {author} {\bibfnamefont {T.}~\bibnamefont
  {Antal}}, \bibinfo {author} {\bibfnamefont {H.}~\bibnamefont {Ohtsuki}},
  \bibinfo {author} {\bibfnamefont {J.}~\bibnamefont {Wakeley}}, \bibinfo
  {author} {\bibfnamefont {P.~D.}\ \bibnamefont {Taylor}},\ and\ \bibinfo
  {author} {\bibfnamefont {M.~A.}\ \bibnamefont {Nowak}},\ }\bibfield  {title}
  {\bibinfo {title} {{Evolution of cooperation by phenotypic similarity}},\
  }\href {https://doi.org/10.1073/pnas.0902528106} {\bibfield  {journal}
  {\bibinfo  {journal} {Proc. Natl. Acad. Sci. U.S.A.}\ }\textbf {\bibinfo
  {volume} {106}},\ \bibinfo {pages} {8597} (\bibinfo {year}
  {2009})}\BibitemShut {NoStop}%
\bibitem [{\citenamefont {Constable}\ \emph {et~al.}(2016)\citenamefont
  {Constable}, \citenamefont {Rogers}, \citenamefont {McKane},\ and\
  \citenamefont {Tarnita}}]{Constable2016}%
  \BibitemOpen
  \bibfield  {author} {\bibinfo {author} {\bibfnamefont {G.~W.}\ \bibnamefont
  {Constable}}, \bibinfo {author} {\bibfnamefont {T.}~\bibnamefont {Rogers}},
  \bibinfo {author} {\bibfnamefont {A.~J.}\ \bibnamefont {McKane}},\ and\
  \bibinfo {author} {\bibfnamefont {C.~E.}\ \bibnamefont {Tarnita}},\
  }\bibfield  {title} {\bibinfo {title} {{Demographic noise can reverse the
  direction of deterministic selection}},\ }\href
  {https://doi.org/10.1073/pnas.1603693113} {\bibfield  {journal} {\bibinfo
  {journal} {Proc. Natl. Acad. Sci. U.S.A.}\ }\textbf {\bibinfo {volume}
  {113}},\ \bibinfo {pages} {E4745} (\bibinfo {year} {2016})}\BibitemShut
  {NoStop}%
\bibitem [{\citenamefont {Chen}\ \emph
  {et~al.}(2015{\natexlab{a}})\citenamefont {Chen}, \citenamefont {Sasaki},
  \citenamefont {Br{\"{a}}nnstr{\"{o}}m},\ and\ \citenamefont
  {Dieckmann}}]{Chen2015inter}%
  \BibitemOpen
  \bibfield  {author} {\bibinfo {author} {\bibfnamefont {X.}~\bibnamefont
  {Chen}}, \bibinfo {author} {\bibfnamefont {T.}~\bibnamefont {Sasaki}},
  \bibinfo {author} {\bibfnamefont {{\AA}.}~\bibnamefont
  {Br{\"{a}}nnstr{\"{o}}m}},\ and\ \bibinfo {author} {\bibfnamefont
  {U.}~\bibnamefont {Dieckmann}},\ }\bibfield  {title} {\bibinfo {title}
  {{First carrot, then stick: How the adaptive hybridization of incentives
  promotes cooperation}},\ }\href {https://doi.org/10.1098/rsif.2014.0935}
  {\bibfield  {journal} {\bibinfo  {journal} {J. R. Soc. Interface}\ }\textbf
  {\bibinfo {volume} {12}},\ \bibinfo {pages} {20140935} (\bibinfo {year}
  {2015}{\natexlab{a}})}\BibitemShut {NoStop}%
\bibitem [{\citenamefont {Posner}(1983)}]{Posner1983}%
  \BibitemOpen
  \bibfield  {author} {\bibinfo {author} {\bibfnamefont {R.~A.}\ \bibnamefont
  {Posner}},\ }\href@noop {} {\emph {\bibinfo {title} {{The economics of
  justice}}}}\ (\bibinfo  {publisher} {Harvard University Press},\ \bibinfo
  {address} {Cambridge},\ \bibinfo {year} {1983})\BibitemShut {NoStop}%
\bibitem [{\citenamefont {Chen}\ \emph
  {et~al.}(2015{\natexlab{b}})\citenamefont {Chen}, \citenamefont {Sasaki},\
  and\ \citenamefont {Perc}}]{Chen2015}%
  \BibitemOpen
  \bibfield  {author} {\bibinfo {author} {\bibfnamefont {X.}~\bibnamefont
  {Chen}}, \bibinfo {author} {\bibfnamefont {T.}~\bibnamefont {Sasaki}},\ and\
  \bibinfo {author} {\bibfnamefont {M.}~\bibnamefont {Perc}},\ }\bibfield
  {title} {\bibinfo {title} {{Evolution of public cooperation in a monitored
  society with implicated punishment and within-group enforcement}},\ }\href
  {https://doi.org/10.1038/srep17050} {\bibfield  {journal} {\bibinfo
  {journal} {Sci. Rep.}\ }\textbf {\bibinfo {volume} {5}},\ \bibinfo {pages}
  {17050} (\bibinfo {year} {2015}{\natexlab{b}})}\BibitemShut {NoStop}%
\bibitem [{\citenamefont {Fu}\ \emph {et~al.}(2008)\citenamefont {Fu},
  \citenamefont {Hauert}, \citenamefont {Nowak},\ and\ \citenamefont
  {Wang}}]{Fu2008}%
  \BibitemOpen
  \bibfield  {author} {\bibinfo {author} {\bibfnamefont {F.}~\bibnamefont
  {Fu}}, \bibinfo {author} {\bibfnamefont {C.}~\bibnamefont {Hauert}}, \bibinfo
  {author} {\bibfnamefont {M.~A.}\ \bibnamefont {Nowak}},\ and\ \bibinfo
  {author} {\bibfnamefont {L.}~\bibnamefont {Wang}},\ }\bibfield  {title}
  {\bibinfo {title} {{Reputation-based partner choice promotes cooperation in
  social networks}},\ }\href {https://doi.org/10.1103/PhysRevE.78.026117}
  {\bibfield  {journal} {\bibinfo  {journal} {Phys. Rev. E}\ }\textbf {\bibinfo
  {volume} {78}},\ \bibinfo {pages} {026117} (\bibinfo {year}
  {2008})}\BibitemShut {NoStop}%
\bibitem [{\citenamefont {Suzuki}\ and\ \citenamefont
  {Akiyama}(2005)}]{Suzuki2005}%
  \BibitemOpen
  \bibfield  {author} {\bibinfo {author} {\bibfnamefont {S.}~\bibnamefont
  {Suzuki}}\ and\ \bibinfo {author} {\bibfnamefont {E.}~\bibnamefont
  {Akiyama}},\ }\bibfield  {title} {\bibinfo {title} {{Reputation and the
  evolution of cooperation in sizable groups}},\ }\href
  {https://doi.org/10.1098/rspb.2005.3072} {\bibfield  {journal} {\bibinfo
  {journal} {Proc. R. Soc. B}\ }\textbf {\bibinfo {volume} {272}},\ \bibinfo
  {pages} {1373} (\bibinfo {year} {2005})}\BibitemShut {NoStop}%
\bibitem [{\citenamefont {{Dos Santos}}\ \emph {et~al.}(2011)\citenamefont
  {{Dos Santos}}, \citenamefont {Rankin},\ and\ \citenamefont
  {Wedekind}}]{DosSantos2011}%
  \BibitemOpen
  \bibfield  {author} {\bibinfo {author} {\bibfnamefont {M.}~\bibnamefont {{Dos
  Santos}}}, \bibinfo {author} {\bibfnamefont {D.~J.}\ \bibnamefont {Rankin}},\
  and\ \bibinfo {author} {\bibfnamefont {C.}~\bibnamefont {Wedekind}},\
  }\bibfield  {title} {\bibinfo {title} {{The evolution of punishment through
  reputation}},\ }\href {https://doi.org/10.1098/rspb.2010.1275} {\bibfield
  {journal} {\bibinfo  {journal} {Proc. R. Soc. B}\ }\textbf {\bibinfo {volume}
  {278}},\ \bibinfo {pages} {371} (\bibinfo {year} {2011})}\BibitemShut
  {NoStop}%
\bibitem [{\citenamefont {Pacheco}\ \emph {et~al.}(2009)\citenamefont
  {Pacheco}, \citenamefont {Santos}, \citenamefont {Souza},\ and\ \citenamefont
  {Skyrms}}]{Pacheco2009}%
  \BibitemOpen
  \bibfield  {author} {\bibinfo {author} {\bibfnamefont {J.~M.}\ \bibnamefont
  {Pacheco}}, \bibinfo {author} {\bibfnamefont {F.~C.}\ \bibnamefont {Santos}},
  \bibinfo {author} {\bibfnamefont {M.~O.}\ \bibnamefont {Souza}},\ and\
  \bibinfo {author} {\bibfnamefont {B.}~\bibnamefont {Skyrms}},\ }\bibfield
  {title} {\bibinfo {title} {{Evolutionary dynamics of collective action in
  N-person stag hunt dilemmas}},\ }\href
  {https://doi.org/10.1098/rspb.2008.1126} {\bibfield  {journal} {\bibinfo
  {journal} {Proc. R. Soc. B}\ }\textbf {\bibinfo {volume} {276}},\ \bibinfo
  {pages} {315} (\bibinfo {year} {2009})}\BibitemShut {NoStop}%
\bibitem [{\citenamefont {Su}\ \emph {et~al.}(2019{\natexlab{b}})\citenamefont
  {Su}, \citenamefont {Zhou},\ and\ \citenamefont {Wang}}]{Su2019plos}%
  \BibitemOpen
  \bibfield  {author} {\bibinfo {author} {\bibfnamefont {Q.}~\bibnamefont
  {Su}}, \bibinfo {author} {\bibfnamefont {L.}~\bibnamefont {Zhou}},\ and\
  \bibinfo {author} {\bibfnamefont {L.}~\bibnamefont {Wang}},\ }\bibfield
  {title} {\bibinfo {title} {{Evolutionary multiplayer games on graphs with
  edge diversity}},\ }\href {https://doi.org/10.1371/journal.pcbi.1006947}
  {\bibfield  {journal} {\bibinfo  {journal} {PLoS Comput. Biol.}\ }\textbf
  {\bibinfo {volume} {15}},\ \bibinfo {pages} {e1006947} (\bibinfo {year}
  {2019}{\natexlab{b}})}\BibitemShut {NoStop}%
\bibitem [{\citenamefont {Li}\ \emph {et~al.}(2016)\citenamefont {Li},
  \citenamefont {Broom}, \citenamefont {Du},\ and\ \citenamefont
  {Wang}}]{Li2016}%
  \BibitemOpen
  \bibfield  {author} {\bibinfo {author} {\bibfnamefont {A.}~\bibnamefont
  {Li}}, \bibinfo {author} {\bibfnamefont {M.}~\bibnamefont {Broom}}, \bibinfo
  {author} {\bibfnamefont {J.}~\bibnamefont {Du}},\ and\ \bibinfo {author}
  {\bibfnamefont {L.}~\bibnamefont {Wang}},\ }\bibfield  {title} {\bibinfo
  {title} {{Evolutionary dynamics of general group interactions in structured
  populations}},\ }\href {https://doi.org/10.1103/PhysRevE.93.022407}
  {\bibfield  {journal} {\bibinfo  {journal} {Phys. Rev. E}\ }\textbf {\bibinfo
  {volume} {93}},\ \bibinfo {pages} {022407} (\bibinfo {year}
  {2016})}\BibitemShut {NoStop}%
\bibitem [{\citenamefont {Su}\ \emph {et~al.}(2019{\natexlab{c}})\citenamefont
  {Su}, \citenamefont {Li}, \citenamefont {Wang},\ and\ \citenamefont
  {Stanley}}]{Su2019prsb}%
  \BibitemOpen
  \bibfield  {author} {\bibinfo {author} {\bibfnamefont {Q.}~\bibnamefont
  {Su}}, \bibinfo {author} {\bibfnamefont {A.}~\bibnamefont {Li}}, \bibinfo
  {author} {\bibfnamefont {L.}~\bibnamefont {Wang}},\ and\ \bibinfo {author}
  {\bibfnamefont {H.~E.}\ \bibnamefont {Stanley}},\ }\bibfield  {title}
  {\bibinfo {title} {{Spatial reciprocity in the evolution of cooperation}},\
  }\href {https://doi.org/10.1098/RSPB.2019.0041} {\bibfield  {journal}
  {\bibinfo  {journal} {Proc. R. Soc. B}\ }\textbf {\bibinfo {volume} {286}},\
  \bibinfo {pages} {20190041} (\bibinfo {year}
  {2019}{\natexlab{c}})}\BibitemShut {NoStop}%
\bibitem [{\citenamefont {Allen}\ and\ \citenamefont
  {McAvoy}(2019)}]{Allen2019}%
  \BibitemOpen
  \bibfield  {author} {\bibinfo {author} {\bibfnamefont {B.}~\bibnamefont
  {Allen}}\ and\ \bibinfo {author} {\bibfnamefont {A.}~\bibnamefont {McAvoy}},\
  }\bibfield  {title} {\bibinfo {title} {{A mathematical formalism for natural
  selection with arbitrary spatial and genetic structure}},\ }\href
  {https://doi.org/10.1007/s00285-018-1305-z} {\bibfield  {journal} {\bibinfo
  {journal} {J. Math. Biol.}\ }\textbf {\bibinfo {volume} {78}},\ \bibinfo
  {pages} {1147} (\bibinfo {year} {2019})}\BibitemShut {NoStop}%
\bibitem [{\citenamefont {Li}\ \emph {et~al.}(2020)\citenamefont {Li},
  \citenamefont {Zhou}, \citenamefont {Su}, \citenamefont {Cornelius},
  \citenamefont {Liu}, \citenamefont {Wang},\ and\ \citenamefont
  {Levin}}]{Li2020}%
  \BibitemOpen
  \bibfield  {author} {\bibinfo {author} {\bibfnamefont {A.}~\bibnamefont
  {Li}}, \bibinfo {author} {\bibfnamefont {L.}~\bibnamefont {Zhou}}, \bibinfo
  {author} {\bibfnamefont {Q.}~\bibnamefont {Su}}, \bibinfo {author}
  {\bibfnamefont {S.~P.}\ \bibnamefont {Cornelius}}, \bibinfo {author}
  {\bibfnamefont {Y.~Y.}\ \bibnamefont {Liu}}, \bibinfo {author} {\bibfnamefont
  {L.}~\bibnamefont {Wang}},\ and\ \bibinfo {author} {\bibfnamefont {S.~A.}\
  \bibnamefont {Levin}},\ }\bibfield  {title} {\bibinfo {title} {{Evolution of
  cooperation on temporal networks}},\ }\href
  {https://doi.org/10.1038/s41467-020-16088-w} {\bibfield  {journal} {\bibinfo
  {journal} {Nat. Commun.}\ }\textbf {\bibinfo {volume} {11}},\ \bibinfo
  {pages} {2259} (\bibinfo {year} {2020})}\BibitemShut {NoStop}%
\end{thebibliography}%
\end{document}